\documentclass[twoside,english,preprint,authoryear,12pt,times,review,a4paper]{elsarticle}
\usepackage[LGR,T1]{fontenc}
\usepackage[latin9]{inputenc}
\usepackage{array}
\usepackage{verbatim}
\usepackage{multirow}
\usepackage{graphicx}
\usepackage{subscript}

\makeatletter

\DeclareRobustCommand{\greektext}{%
  \fontencoding{LGR}\selectfont\def\encodingdefault{LGR}}
\DeclareRobustCommand{\textgreek}[1]{\leavevmode{\greektext #1}}
\ProvideTextCommand{\~}{LGR}[1]{\char126#1}

\providecommand{\tabularnewline}{\\}

\journal{Icarus}

\usepackage{lineno}
\usepackage{longtable}
\usepackage{dcolumn}

\usepackage[font=scriptsize,labelfont=bf]{caption}
\usepackage{adjustbox}

\makeatother

\usepackage{babel}
\begin{document}

\begin{frontmatter}{}

\title{Colour and Tropospheric Cloud Structure of Jupiter from MUSE/VLT:
Retrieving a Universal Chromophore}

\author[oxford,latmos]{Ashwin S. Braude\corref{cor1}}

\ead{ashwin.braude@latmos.ipsl.fr}

\author[oxford]{Patrick G. J. Irwin}

\author[jpl]{Glenn S. Orton}

\author[leicester]{Leigh N. Fletcher}

\cortext[cor1]{Corresponding author}

\address[oxford]{Department of Physics, University of Oxford, Parks Rd, Oxford, OX1
3PU, UK.}

\address[latmos]{Laboratoire Atmosph\`{e}res, Milieux, Observations Spatiales (LATMOS),
UVSQ Universit\'{e} Paris-Saclay, Sorbonne Universit\'{e}, CNRS, Paris, France
}

\address[jpl]{Jet Propulsion Laboratory, California Institute of Technology, 4800
Oak Grove Drive, Pasadena, CA 91109, USA.}

\address[leicester]{Department of Physics \& Astronomy, University of Leicester, University
Road, Leicester, LE1 7RH, UK.}
\begin{abstract}
Recent work by Sromovsky et al. (2017, Icarus 291, 232-244) suggested
that all red colour in Jupiter's atmosphere could be explained by
a single colour-carrying compound, a so-called `universal chromophore'.
We tested this hypothesis on ground-based spectroscopic observations
in the visible and near-infrared (480-930 nm) from the VLT/MUSE instrument
between 2014 and 2018, retrieving a chromophore absorption spectrum
directly from the North Equatorial Belt, and applying it to model
spatial variations in colour, tropospheric cloud and haze structure
on Jupiter. We found that we could model both the belts and the Great
Red Spot of Jupiter using the same chromophore compound, but that
this chromophore must exhibit a steeper blue-absorption gradient than
the proposed chromophore of Carlson et al. (2016, Icarus 274, 106\textendash 115).
We retrieved this chromophore to be located no deeper than $0.2\pm0.1$
bars in the Great Red Spot and $0.7\pm0.1$ bars elsewhere on Jupiter.
However, we also identified some spectral variability between 510
nm and 540 nm that could not be accounted for by a universal chromophore.
In addition, we retrieved a thick, global cloud layer at $1.4\pm0.3$
bars that was relatively spatially invariant in altitude across Jupiter.
We found that this cloud layer was best characterised by a real refractive
index close to that of ammonia ice in the belts and the Great Red
Spot, and poorly characterised by a real refractive index of 1.6 or
greater. This may be the result of ammonia cloud at higher altitude
obscuring a deeper cloud layer of unknown composition.
\end{abstract}

\end{frontmatter}{}


\section{Introduction}

Spatial and temporal variation in red colour on Jupiter is thought
to be due to the presence of red colour-carrying compounds (`chromophores')
whose origin, composition and altitude remain unknown. A number of
factors contribute to the difficulty of identifying Jovian chromophore
composition. These include the apparent absence of characteristic
absorption bands other than a broad absorption feature at blue wavelengths,
a relative lack of laboratory studies performed under realistic Jovian
conditions, and comprehensive spectral data of Jupiter at the required
wavelength range and resolution. Historically, most scientific studies
of colour on Jupiter have relied on a restricted number of wavelength
filters (eg. \citealp{Owen1981,Thompson1990,simon2001color,strycker2011jovian,Ordonez-Etxeberria2016}),
but various effects (such as chromophore composition, altitude and
particle size) can alter the shape of Jovian spectra at short wavelengths,
which can only be decoupled from each other through the observation
of Jovian spectra at greater spectral resolution than filter imaging
can provide. The last global sets of hyperspectral data of Jupiter
in the visible wavelength range were obtained from the Cassini/VIMS-V
instrument \citep{Brown2004,coradini2004}, which observed Jupiter
at wavelengths between 350 nm and 1100 nm during its brief flyby in
2000/2001. \citet{Sromovsky2017} used these data to propose the idea
of a `universal chromophore' for Jupiter, in which all red colour
on Jupiter originated from the same source compound and reaction process.
This compound was based on the red substance obtained by \citet{carlson2016chromophores}
in the laboratory through the reaction of photolysed ammonia with
acetylene, consisting of a range of complex organic compounds that
absorb strongly at blue wavelengths due to molecular $\pi\rightarrow\pi^{*}$
bond transitions \citep{Nassau1983}. \citet{baines2019} found that
this chromophore was most likely located in a narrow layer in the
upper troposphere just above the level of the ammonia clouds (the
so-called \emph{`Cr\`{e}me Br\^{u}l\'{e}e} model'). However, elements of Jupiter's
appearance have changed considerably since the Cassini flyby. We note
in particular the dramatic reddening of the Great Red Spot (GRS) that
has accompanied its progressive shrinking in size \citep{simon2014dramatic,Simon2018},
as well as progressive cycles of colour changes in the southern North
Temperate Belt (NTB) as reviewed by \citet{Rogers1995} and \citet{Fletcher2017}.
These provide additional opportunities for constraint of the origin
and altitude of chromophores in Jupiter's atmosphere that were not
previously available.

This paper makes use of global, hyperspectral data of Jupiter obtained
at visible and near-infrared wavelengths from the ground-based VLT/MUSE
instrument (described in section 2). Uniquely, these data include
spectra both of the GRS in its current deep red state and of the southern
NTB when it was at its reddest in early 2017. Using a radiative-transfer
model (section 3), and building on the preliminary results presented
in \citeauthor{Irwin2018} \citeyearpar{Irwin2018,2019Icar..321..572I},
we retrieve a range of possible chromophore refractive index solutions
from limb-darkening analysis of the North Equatorial Belt (NEB), given
different prior particle-size distributions, and fit them to a spectrum
of the GRS in order to show that a chromophore solution that is universal
to all red regions of Jupiter is possible (section 4). We then apply
this chromophore solution more broadly to model spatial variations
in colour and tropospheric cloud structure (section 5), showing that
there is some spectral variability between 0.51 \textgreek{m}m and
0.54 \textgreek{m}m that cannot be explained by a universal chromophore
alone. All our findings are summarised in section 6.

\section{Data}

\subsection{Description of VLT/MUSE observations}

The Multi-Unit Spectroscopic Explorer (MUSE), mounted on the European
Southern Observatory's Very Large Telescope (VLT), is an integral-field
spectrograph \citep{bacon2010muse}, whose spectral range nominally
covers wavelengths from 480 nm to 930 nm (in practice closer to 476
to 933 nm), with radiance values sampled at 0.125 nm intervals (R\textasciitilde 1770-3590).
Even at solar opposition, Jupiter's entire visible hemisphere fits
neatly into MUSE's 64'' x 64'' field-of-view, which is subdivided
into individual 2-D spatial pixels of size 0.2'' x 0.2'' (subtending
a great circle of approximately 640 km at nadir when Jupiter is at
5.4 AU from the Sun), each of which contain a single, invertible spectrum
of Jupiter. To date, MUSE is the only instrument that can obtain data
such as this within an exposure time of a fraction of a second. Its
observations therefore play a crucial part in the ground support of
NASA's Juno mission currently in orbit around Jupiter (although we
do not include any analysis of Juno data in this paper), which has
no instruments on board that can observe Jupiter at the required wavelength
range (aside from JunoCam \citep{Hansen2014}, a public outreach camera
with low SNR) and which has a narrow spatial footprint.

\begin{figure}
\includegraphics[width=1\textwidth,height=0.9\textheight,keepaspectratio]{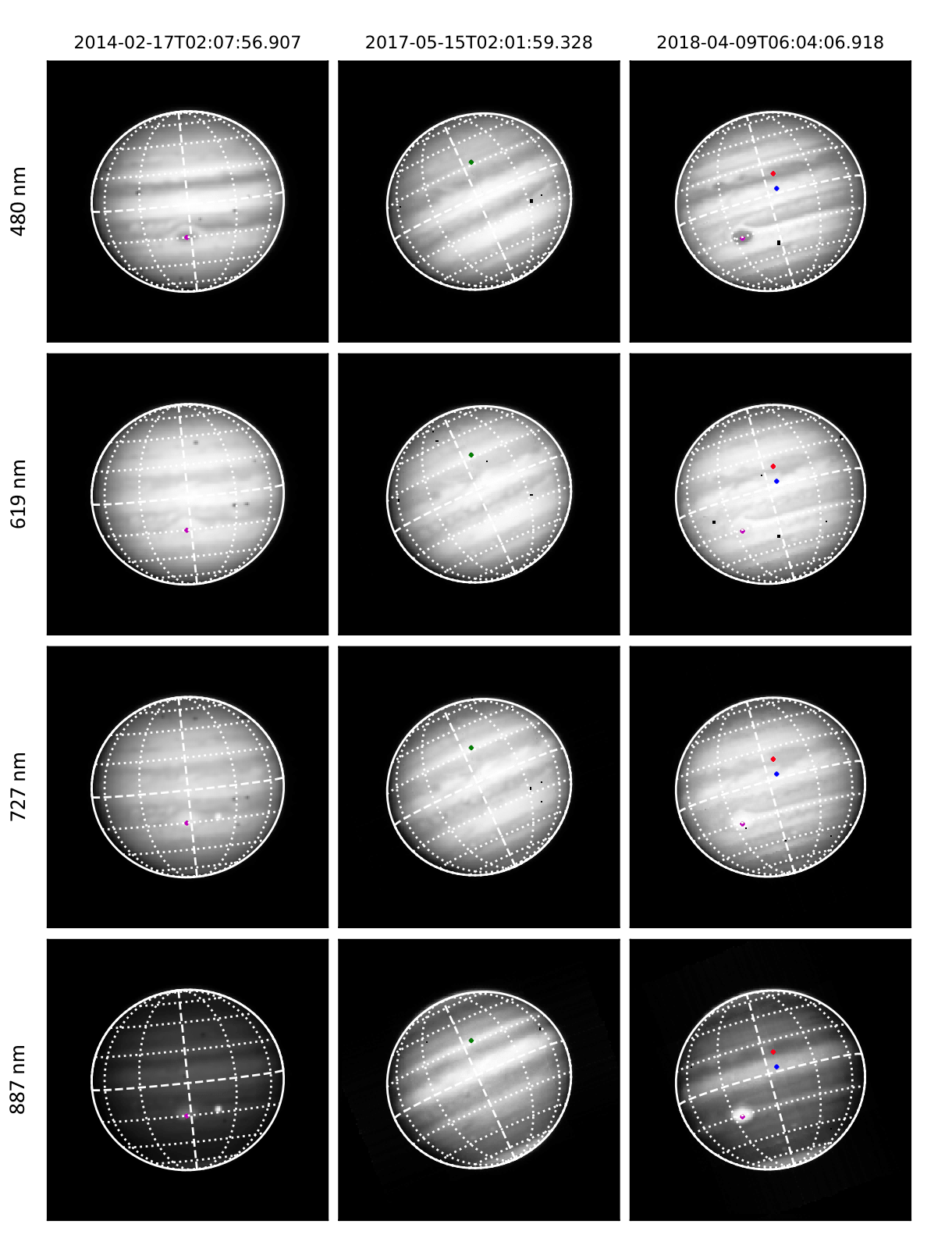}

\caption{An image of each of the three MUSE observations of Jupiter at four
sample wavelengths: 480 nm (sensitive to variations in chromophore),
619 nm (sensitive to variations in the deepest visible cloud layers,
around 1-2 bars), 727 nm (sensitive to variations in upper tropospheric
haze around 0.3-0.5 bars) and 887 nm (sensitive to the highest observable
haze layers in the MUSE wavelength range, around 0.2-0.3 bars). In
each case, the thick white line indicates the calculated terminator
of Jupiter through ellipsoid limb fitting, the white dashed lines
indicate the equator and central meridian (refer to Table \ref{listofobs}
for the System III longitude in each case) and the white dotted lines
indicate lines of planetocentric latitude (in $20^{o}$ intervals)
and longitude (in $30^{o}$ intervals from the central meridian).
Locations of sample spectra from the EZ (in blue), the NEB (in red),
the NTBs (in green) and the GRS (in purple) are marked in each image.
The bright white spot that is visible in the 2014 image around 20$^{o}$
eastwards of the GRS is Europa.}

\label{obsimages}
\end{figure}

A number of sets of observations of Jupiter were obtained by the MUSE
instrument between 2014 and 2018. When combined, these datasets provide
a unique view of temporal changes in the visible appearance of Jupiter
before and during the Juno mission that are largely beyond the scope
of this paper. In order simply to focus on the potential applicability
of a universal chromophore, we instead primarily use a single observation
of Jupiter in April 2018 (timestamp 2018-04-09T06:04:06.918). This
observation was obtained under the best observing conditions (with
seeing of just 0.38'' and an airmass of close to 1) of all the datasets,
and also consisted of an observation of the GRS close to nadir. For
comparison, we have also included an observation from February 2014
(2014-02-17T02:07:56.907), which contained an observation of the GRS
before its current, shrunken state, as well as one from May 2017 (2017-05-15T02:01:59.328)
when the southern NTB was at its reddest following an upheaval in
the region in the previous year \citep{Sanchez-Lavega2017}. These
observations are listed in Table \ref{listofobs} and shown individually
at sample wavelengths in Figure \ref{obsimages}. Given our lack of
knowledge of the uncertainty in the PSF, we have avoided performing
retrievals close to Jupiter's terminator where the surrounding sky
would have provided a non-negligible contribution to the observed
spectrum. On top of these MUSE data, we also make use of a single
ready-calibrated Cassini/VIMS spectrum of the GRS obtained in 2000,
as provided in the supplementary material of \citet{carlson2016chromophores},
to provide broader context to temporal changes in spectra of the GRS.
We do not describe the calibration procedure of this latter observation
here, instead we refer the reader to \citet{baines2019} for a more
detailed explanation.

\subsection{Calibration and post-processing}

Calibration and reduction of each dataset was performed using the
standard ESOREX pipeline \citep{weilbacher2014}. %
Following calibration, the spectral image cubes were projected manually
using ellipsoid limb fitting and the associated navigational metadata
calculated. As described in \citet{Irwin2018}, the resulting spectral
image cubes were then smoothed using a triangular instrument function
of FWHM = 2 nm (R\textasciitilde 200) akin to that of the IRTF/SpeX
instrument \citep{rayner2003spex}, which was then sampled at 1 nm
resolution. This was in order to be compatible with the methane band
data of \citet{Karkoschka2010} and the reference solar spectrum \citep{chance2010improved},
for which higher-resolution data is not available at the required
wavelength range, as well as to save on computational time. Poor flat-fielding
in the 2014-02-17T02:07:56.907 dataset resulted in visible striping
artefacts which were smoothed over by convolving the data with a Gaussian
filter of FWHM = 0.3 arcsecs, thereby resulting in some loss of spatial
resolution compared with the other two datasets.

\begin{figure}
\includegraphics[width=1\textwidth]{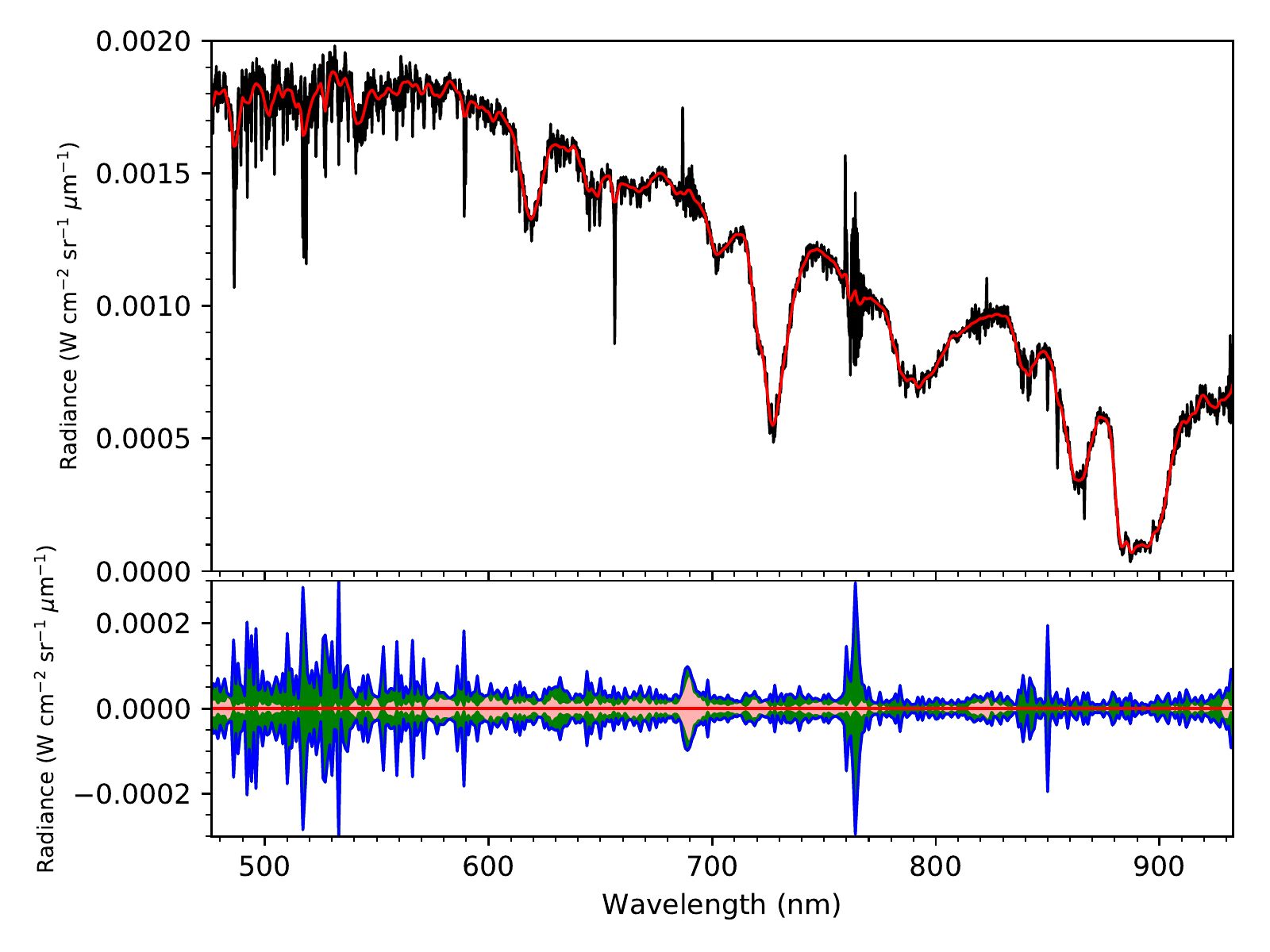}

\caption{Top panel: A sample spectrum of the EZ (black) at native resolution
and (red) following smoothing to 1 nm resolution. Bottom panel: spectral
uncertainties associated with the smoothed EZ spectrum. In light red
is the spectral uncertainty contribution from the errors calculated
by the ESOREX pipeline following the smoothing process, taking into
account spectral correlation. However, this is seen to underestimate
the spectral error at certain wavelengths, notably in regions of telluric
noise (e.g. 760 nm) and in regions where a large number of solar spectral
lines are present (e.g. 500-540 nm). For this reason, we add an extra
bias term (in green) to compensate for oversmoothing of the spectrum.
On top of this we include an additional 1\% uncertainty (in blue)
to account for forward modelling error.}

\label{errors}
\end{figure}

Spectral error was calculated automatically using the standard pipeline,
with the main sources primarily being telluric absorption in the Earth's
atmosphere and interpolation over stellar lines in the standard spectrum,
with smaller contributions from the calculation of spectral response
and readout noise. Smoothing the spectral image cubes to SpeX resolution
was seen to reduce random error to negligible values by comparison.
However, it also introduced additional sources of systematic uncertainty
due to spectral correlation, as well as systematic `bias' errors from
oversmoothing over regions with a high density of solar spectral and
telluric lines, a consequence of the `bias-variance dilemma' \citep{geman1992neural}.
This bias was quantified simply by taking the difference in radiance
at a given wavelength (at SpeX resolution) before and after the smoothing
process. An additional systematic error term was also added onto the
total spectral error term, in order to encompass various sources of
uncertainty in the forward model that are generally difficult to quantify.
These include a) a combination of experimental uncertainties in the
measurement of the reference gas absorption data, b) uncertainties
that result from the use of the correlated-k approximation instead
of the line-by-line method, and c) uncertainties that result from
other approximations made in the radiative-transfer equation. We found
that adding an additional 1\% spectral error of the measured flux
at each wavelength was sufficient to obtain values of the normalised
goodness of fit, $\chi^{2}/n$, at or below 1 in the majority of Jovian
spectra. The contribution of all these factors to the total spectral
uncertainty is shown in Figure \ref{errors}.

We used two main tests to verify the photometric calibration of each
dataset. The first was to disc-average each of the three datasets
and cross-calibrate them with the two disc-averaged spectra of \citet{karkoschka1994spectrophotometry}
and \citet{karkoschka1998methane}, respectively calculated from observations
of Jupiter in 1993 and 1995, with the latter considered to be of superior
quality than the former due to better observing conditions. In the
case of the MUSE datasets, the radiance values were converted to I/F
using the extraterrestrial solar spectrum of \citet{chance2010improved}.
However, Jupiter's appearance changes over both time and longitude,
and so another external source of data, obtained closer in time to
the MUSE observations, is required to fully verify the calibration.
For this reason, we make use of data from the HST/WFC3 instrument,
obtained as part of the OPAL programme \citep{Simon2015}. This data
includes two sets of fully-calibrated and projected global observations
of Jupiter per year, between 2015 and 2018, as viewed in five wavelength
filters that overlap with the MUSE wavelength range (F502N, F547M,
F631N, F658N and FQ889N, each named according to the central wavelength
in nanometres). Reliable cross-calibration between the MUSE and HST/WFC3
datasets is, however, still complicated by both longitudinal variation
and the fact that an empirical Minnaert correction has been applied
to the HST/WFC3 datasets which increases the disc-averaged I/F. When
we multiply the MUSE datasets by each of the five filter functions
and then perform our own empirical Minnaert correction on them, we
find that even very small uncertainties in the Minnaert coefficients
can lead to edge effects at viewing zenith angles close to 90$^{o}$
that can wildly distort the resulting disc-average. To minimise the
effect of the uncertainty of the Minnaert correction on the cross-calibration,
we simply average each MUSE dataset, following filtering and Minnaert
correction, within 60$^{o}$ of latitude either side of the equator
and within $10^{o}$ either side of the sub-observer longitude (a
`swath-average'), and then do the same with the HST/WFC3 datasets
obtained closest in time and averaged around the same sub-observer
longitude.

\begin{figure}
\includegraphics[width=1\columnwidth]{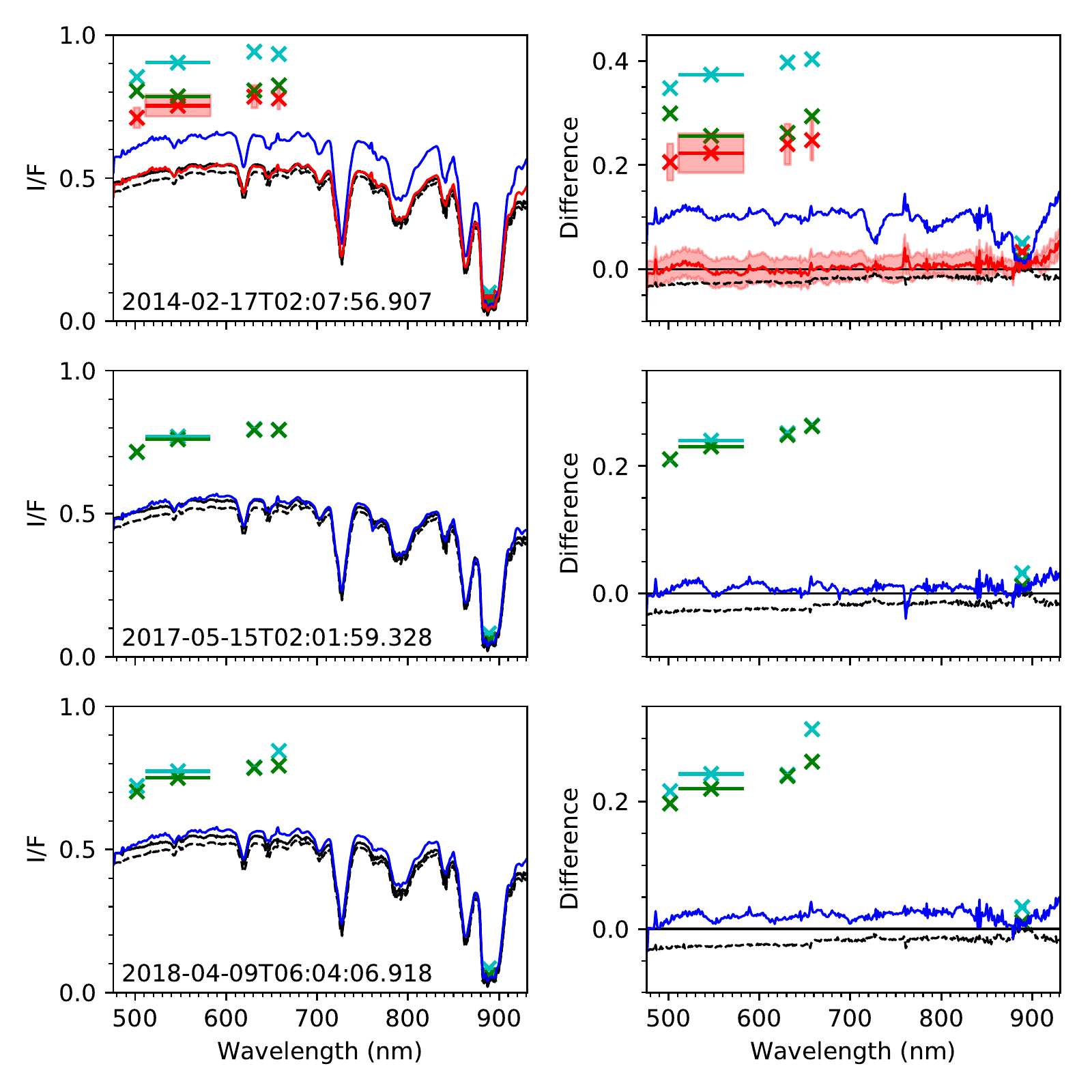}\caption{\label{fig:calib}Verification of calibration of MUSE I/F values against
both the disc-averaged spectra of \citet{karkoschka1994spectrophotometry}
(black, dashed) and \citet{karkoschka1998methane} (black, solid),
and against global HST/WFC3 observations obtained closest in time
(from top to bottom: 19th of January 2015, 3rd of April 2017, 17th
of April 2018) to the corresponding MUSE observation as part of the
OPAL programme \citep{Simon2015}. The MUSE observation timestamp
in each case is given in the bottom-left corner of each plot, and
the disc-average of each MUSE dataset shown in dark blue. For cross-calibration
with the HST/WFC3 observations (green crosses), each MUSE observation
was first multiplied by five UVIS2 HST/WFC3 wavelength filter functions
(F502N, F547M, F631N, F658N and FQ889N) and a Minnaert correction
applied through empirical calculation of limb-darkening over single
zonal swaths. The latitudinal average was then calculated as stated
in the text, and is shown as light blue crosses on the diagram (the
horizontal lines indicate the FWHM of each of the HST filter functions).
We show that the calibration of the MUSE data from 2017 and 2018 is
mostly in keeping with both the \citet{karkoschka1998methane} data
and the HST/WFC3 data, both in terms of the shape of the spectrum
and in the absolute calculation of I/F. However, the MUSE data from
2014 has to be scaled downwards by 20\% over all wavelengths (the
scaled disc-averaged spectrum shown in red, with 5\% uncertainties
shaded) in order to be within 5\% of the Karkoschka and HST/WFC3 data.}
\label{calib}
\end{figure}

As we show in Figure \ref{fig:calib}, the general shape of each of
the three disc-averaged MUSE spectra correspond well with those of
\citet{karkoschka1994spectrophotometry} and \citet{karkoschka1998methane},
and we observe a particularly low systematic offset in I/F between
the \citet{karkoschka1998methane} disc-averaged spectrum and the
2017-05-15T02:01:59.328 and 2018-04-09T06:04:06.918 disc-averaged
spectra respectively. We also observe the swath-averages of the two
MUSE datasets to be within 5\% of those observed by the OPAL programme
on the 3rd of April 2017 and the 17th of April 2018 respectively,
with the only major discrepancy being at the F658N filter in the 2018
datasets, which is a part of the spectrum that is dominated by a solar
spectral line and for which there is therefore some uncertainty in
the calculation of I/F. However, we observe a far greater systematic
offset between the 2014-02-17T02:07:56.907 disc-averaged MUSE spectra
and both the Karkoschka spectra and the HST/WFC3 latitudinal average.
This discrepancy is only partly explicable by the considerable time
that elapsed between the obtention of the 2014-02-17T02:07:56.907
MUSE dataset and the obtention of the first set of OPAL data on the
19th of January 2015, since \citet{Mendikoa2017} only find a major
decrease in I/F in the North Tropical Zone during this time, which
is not enough to fully explain the offset in the latitudinal average.
In addition, this offset in the MUSE data results in I/F values in
the EZ consistently much greater than 1, which is unphysical and results
in atmospheric retrievals that do not converge to a proper solution.
We find that we require a systematic decrease in I/F of 20\% over
all wavelengths in order to have I/F values consistently less than
1 over the whole of Jupiter in this particular dataset. This generally
results in the MUSE spectra being correct to within 5\% of both the
Karkoschka and HST/WFC3 data, with some discrepancy in the F502N filter
possibly due to changes in the GRS. Nonetheless, the relatively poor
calibration of the 2014 data needs to be taken into account when interpreting
our atmospheric retrievals, particularly of tropospheric cloud opacity
which is most affected by systematic offsets in I/F.

\section{Model description}

\subsection{Reference atmosphere}

We made use of the NEMESIS radiative-transfer and retrieval algorithm
\citep{irwin2008nemesis} in order to model all our MUSE spectra.
This works by splitting a reference Jovian atmosphere into a number
of discrete homogeneous layers (we use 39 in this analysis), and solving
the radiative-transfer equation in each layer, taking into account
multiple scattering of aerosols, according to the doubling-adding
method \citep{plass1973matrix}, with the state vector iteratively
adjusted according to a Levenberg-Marquardt scheme \citep{rodgers2000inverse}
in order to provide the best fit to the observed spectra through the
optimal estimation method.

The reference atmosphere covered pressure values from a deep limit
of 10 bars (90km below the 1 bar reference level) to a high-altitude
limit of 1 mbar (150km above the 1 bar level); both altitudes lie
far outside MUSE's sensitivity range. The 39 homogeneous layers of
the reference atmosphere were spaced closer together in the regions
of greatest vertical sensitivity (peaking at intervals of approximately
a tenth of a pressure scale height between 0.1 and 2 bars) but further
apart in regions of lesser vertical sensitivity. The temperature-pressure
profile from the high stratosphere down to 0.8 bars was obtained by
averaging a range of temperature profiles inverted from Cassini/CIRS
observations of Jupiter acquired during the 2000 flyby \citep{fletcher2009phosphine},
which was then extrapolated to deeper altitudes using a dry adiabat.
We neglected spatial variations in temperature, for which no information
could be retrieved from visible and near-infrared spectra. Deep volume
mixing ratios of H\textsubscript{2}, He and CH\textsubscript{4}
were set to 0.86, 0.134 and $1.8\times10^{-3}$ respectively \citep{niemann1998composition,zahn1998helium},
while our prior gaseous ammonia profile was based on that of \citet{fletcher2009phosphine}.

\subsection{Gas absorption models}

By far the greatest sources of gas absorption in this wavelength regime
are methane and ammonia. Accurate methane line lists are lacking at
these wavelengths, and so methane absorption data was modelled according
to \citet{Karkoschka2010} from a combination of measurements in the
laboratory and methane transmission measurements through Titan's atmosphere
obtained from the Huygens probe. Each absorption band was approximated
with a Goody-Voigt band model \citep{goody1995atmospheric}, and then
k-tables computed through exponential line fitting \citep{Irwin2018}.
Ammonia absorption data were obtained from the new \citet{coles2018improved}
line list, with k-distributions computed as in \citet{2019Icar..321..572I}.
The presence of hydrogen gas in Jupiter also induces substantial absorption
around 810-830 nm, primarily due to collision-induced absorption (CIA)
of H\textsubscript{2} - He and H\textsubscript{2} - H\textsubscript{2},
which we modelled according to \citet{borysow1989h2hecia} and \citet{borysow2000cia}
respectively, as well as some additional, very narrow quadrupole absorption
lines for which we used the line lists found in the HITRAN database
\citep{Rothman2013}. We neglected the modelling of higher-order CIA
and quadrupole absorption lines between 620 nm and 640 nm, which are
not discernable in our MUSE spectra and for which accurate line data
are not present in the literature. We make use of the correlated-k
approximation \citep{goody1989correlated} when modelling methane-,
ammonia- and quadrupole H\textsubscript{2} gas absorption, whose
k-distributions are then combined according to the overlapping line
approximation \citep{Lacis1991} in order to save computational time.
Rayleigh scattering cross-section computation codes were obtained
from Sromovsky (personal communication) and modelled using standard
theory (eg. \citealp{goody1995atmospheric}), as explained in detail
in \citet{irwinneptune}. The reference extraterrestrial solar spectrum,
used to convert radiance values into I/F, was obtained from \citet{chance2010improved}.
All our k-tables and our solar spectrum were smoothed using a triangular
instrument function with a FWHM of 2 nm in order to be compatible
with the MUSE spectra (or using a Gaussian instrument function with
a FWHM of 7 nm when performing retrievals on VIMS spectra).

\subsection{Cloud and chromophore model\label{subsec:Cloud-and-chromophore}}

\begin{figure}
\includegraphics[width=1\columnwidth]{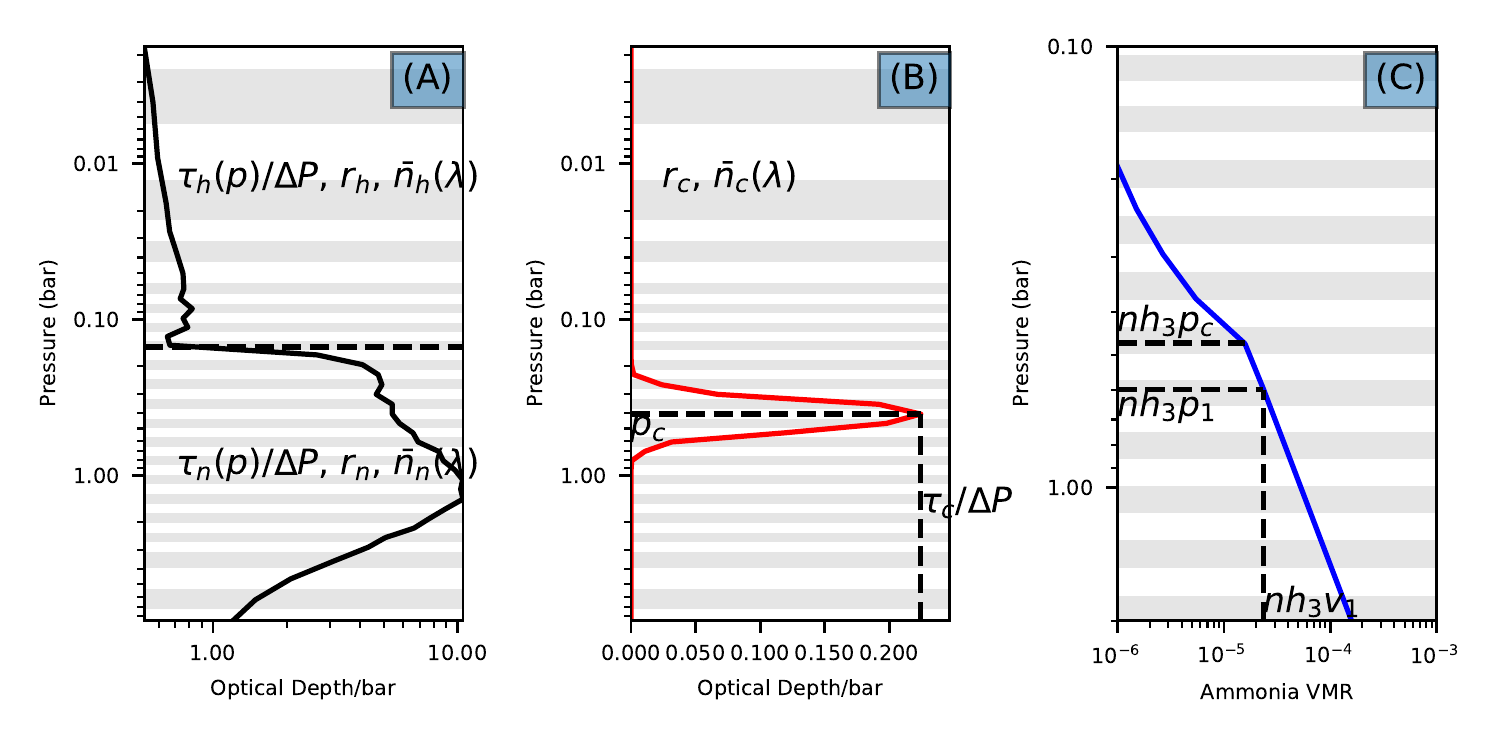}

\caption{\label{fig:cloudmodel}Illustration of the cloud model used in this
work, showing a retrieved profile from spectral limb-darkening of
the NEB when $n_{n}=1.42$, $r_{n}=1\mu m$ and $r_{c}=0.05\mu m$.
Plot (A) shows conservatively-scattering aerosol retrieved continuously
over the entire vertical extent of the atmosphere. For pressure levels
deeper than 0.15 bars, the aerosol is assumed to be `cloud', characterised
by an optical depth per bar (that is, the optical depth divided by
the pressure interval $\Delta p$ of the homogeneous layer in question)
as a function of pressure $\tau_{n}(p)/\Delta p$ (with correlation
length 1.5), size distribution following a Gamma distribution of effective
radius $r_{n}$ and variance 0.05, and complex refractive index as
a function of wavelength $\bar{n}_{n}(\lambda)=n_{n}(\lambda)+ik_{n}(\lambda)$,
where $k_{n}(\lambda)$ is assumed close to zero for all wavelengths
and is parametrised with a correlation length of 0.1. For pressure
levels shallower than 0.15 bars, the aerosol is assumed to be `haze'
and parametrised analogously using the variables $\tau_{h}(p)/\Delta p$,
$r_{h}$ and $n_{h}(\lambda)$. Plot (B) shows chromophore abundance
retrieved as a Gaussian profile, of peak optical depth per bar $\tau_{c}/\Delta p$
at pressure level $p_{c}$ independently retrieved and with a FWHM
of a quarter of a pressure scale height, and with $r_{c}$ and $\bar{n}_{c}(\lambda)=n_{c}(\lambda)+ik_{c}(\lambda)$
parametrised analogously to conservatively-scattering aerosol. Plot
(C) shows the gaseous ammonia profile with two parameters independently
retrieved: a volume mixing ratio value $nh_{3}v_{1}$ at a reference
pressure level $nh_{3}p_{1}=0.6$ bars, and an exponentional fractional
scale height $fsh$ retrieved between the bottom of the atmosphere
and the condensation level $nh_{3}p_{c}$, pressures shallower than
which ammonia is assumed saturated in the atmosphere (note that plot
C is on a different pressure scale than plots A and B as we are only
sensitive to ammonia abundances over a very small vertical range and
the abundance declines very rapidly with height above the saturation
level). We have also shaded in the homogeneous layers of the reference
atmosphere in alternating white and grey horizontal bands.}
\end{figure}

In this analysis we wished to characterise three main populations
of aerosols according to the information contained within the MUSE
spectra: a thick tropospheric cloud layer of large conservatively-scattering
particles, a high-altitude haze layer of small conservatively scattering
particles, and a layer of chromophore particles which we assumed were
responsible for all the blue-absorption seen on Jupiter. Throughout
the rest of this paper we will therefore refer to these particle populations
as `cloud', `haze' and `chromophore' respectively. We modelled both
the cloud and haze layers as continuous abundance profiles in order
to make the best use of the vertical resolution provided by the MUSE
spectra, the haze profile occupying the upper half of the atmosphere
and the cloud profile occupying the lower half, separated by a narrow
boundary layer at 0.15 bars (a pressure level which we found provided
the best fit to spectral limb-darkening in both the EZ and NEB at
890 nm, although in practice we are only weakly sensitive to pressure
levels below this when observing at single geometries) as shown in
Figure \ref{fig:cloudmodel}. Both profiles were sampled at intervals
of half a pressure scale height where vertical sensitivity was greatest
(between 0.1 and 2 bars) and with more minimal sampling where vertical
sensitivity was poorer.

By contrast, we modelled the chromophore layer using a Gaussian profile,
with optical depth as a function of pressure $\tau_{c}(P)$ parametrised
according to the following equation:

\[
\tau_{c}(P)=\tau_{c0}\exp\left(-\left(\frac{\ln(P/P_{c})}{\Delta_{c}}\right)^{2}\right)
\]
where $\tau_{c0}$ is the peak optical depth value and $P_{c}$ the
pressure level at which this occurs, which are both allowed to vary
freely. Little information can be derived on the chromophore vertical
extent $\Delta_{c}$ from the MUSE spectra, and so we chose to use
a value $\Delta_{c}=0.25$. This was seen to be narrow enough to facilitate
reliable comparison with the so-called \emph{Cr\`{e}me Br\^{u}l\'{e}e} model of
\citet{baines2019}, which modelled a chromophore layer with very
narrow vertical extent, but wide enough for $P_{c}$ to be reliably
retrieved given the vertical resolution of the reference atmosphere.
The choice of FWHM has no measurable influence on the chromophore
imaginary refractive index, and only a weak influence on the retrieved
chromophore pressure level. We chose this approach over the so-called
\emph{Cr\`{e}me Br\^{u}l\'{e}e} model of \citet{baines2019} for several reasons:
firstly, it prevented the retrieval of chromophore altitude from interfering
with retrievals of cloud and haze, and secondly, it allowed for the
modelling of an upper-tropospheric haze layer around 0.2-0.3 bars%
{} to fit the strong 890 nm methane absorption feature that could be
decoupled from tropospheric cloud located at higher pressures. At
this wavelength range, there is very little sensitivity to stratospheric
haze (p < 0.15 bars) from single-geometry observations, except at
the poles which are observed at high viewing zenith angle and where
the haze is particularly opaque. To constrain stratospheric haze opacity
close to nadir, we require information at wavelengths below approximately
420 nm \citep{fry2018poster}. For a more detailed breakdown of the
more stylistic differences between the \emph{Cr\`{e}me Br\^{u}l\'{e}e} model and
our own, we refer the reader to the thesis of \citet{mythesis}. For
reference, we will also compare some of the retrievals that we have
conducted using this model with those using the \emph{Cr\`{e}me Br\^{u}l\'{e}e
}model where stated, using a modelling approach as close to that of
\citet{Sromovsky2017} as we can replicate.

\begin{figure}
\includegraphics[width=0.5\textwidth,height=0.5\textheight,keepaspectratio]{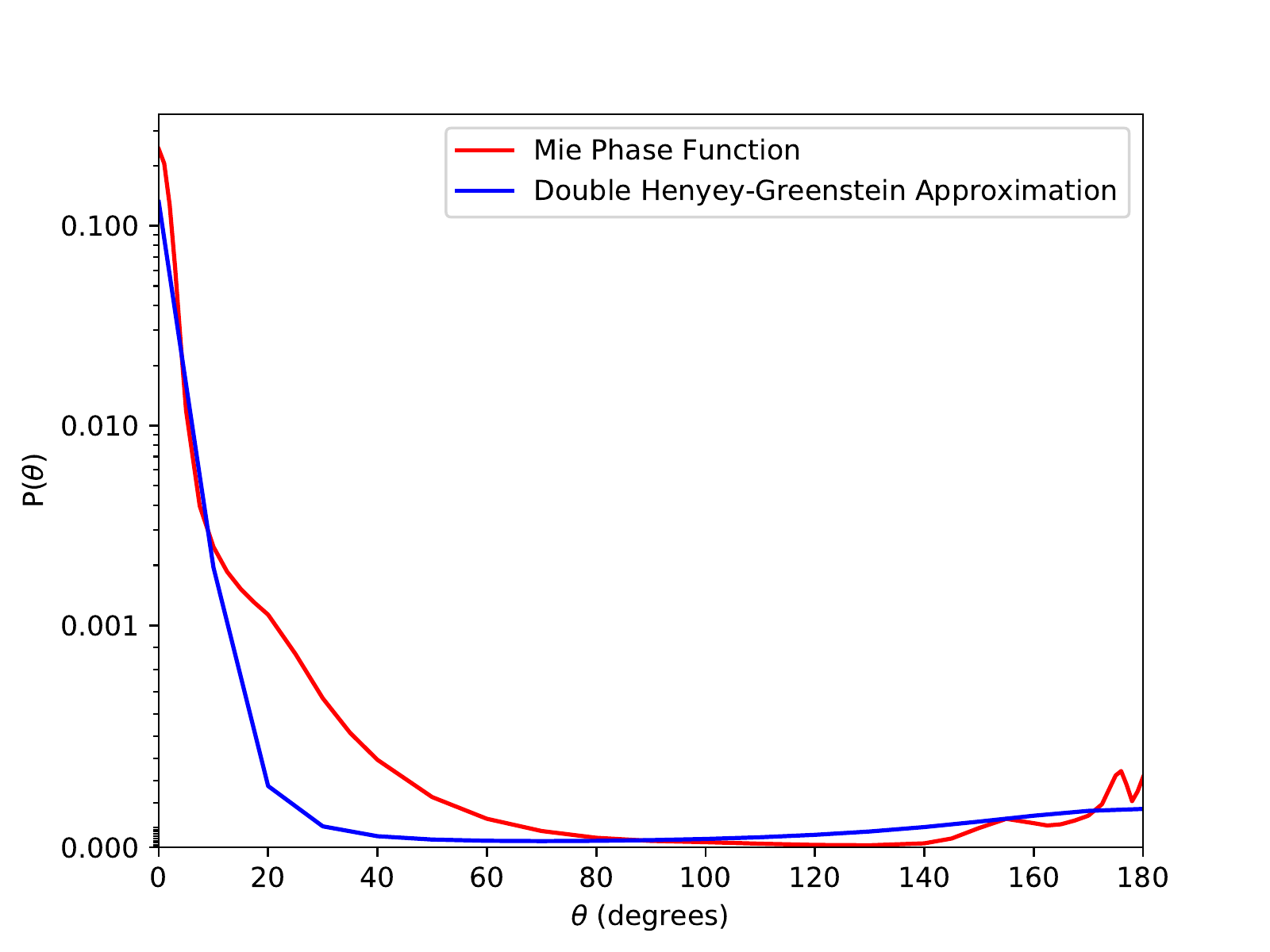}

\caption{Example normalised Mie phase function (for a conservatively-scattering
particle of radius 4\textgreek{m}m) at a wavelength of 700 nm, compared
to the equivalent `smoothed' phase function computed using the double
Henyey-Greenstein approximation.}

\label{phasefuncs}
\end{figure}

Although cloud and haze particles in Jupiter's atmosphere are solid
and therefore non-spherical, we assume that a large ensemble of non-spherical
particles can be modelled as spherical Mie scatterers using the code
of \citet{Dave1968}. We then approximate the resulting Mie phase
functions using a double Henyey-Greenstein phase function \citep{henyey1941diffuse},
as calculated using a Levenberg-Marquardt scheme that minimises the
least-squares deviation from the Mie phase function, in order to smooth
over features of the Mie phase functions that are unique to spherical
particles. A comparison of these two phase functions is shown in Figure
\ref{phasefuncs}.

\section{Retrieving a universal chromophore}

\subsection{Limb-darkening analysis\label{subsec:Limb-darkening-analysis}}

\begin{figure}
\includegraphics[width=1\textwidth,height=0.9\textheight,keepaspectratio]{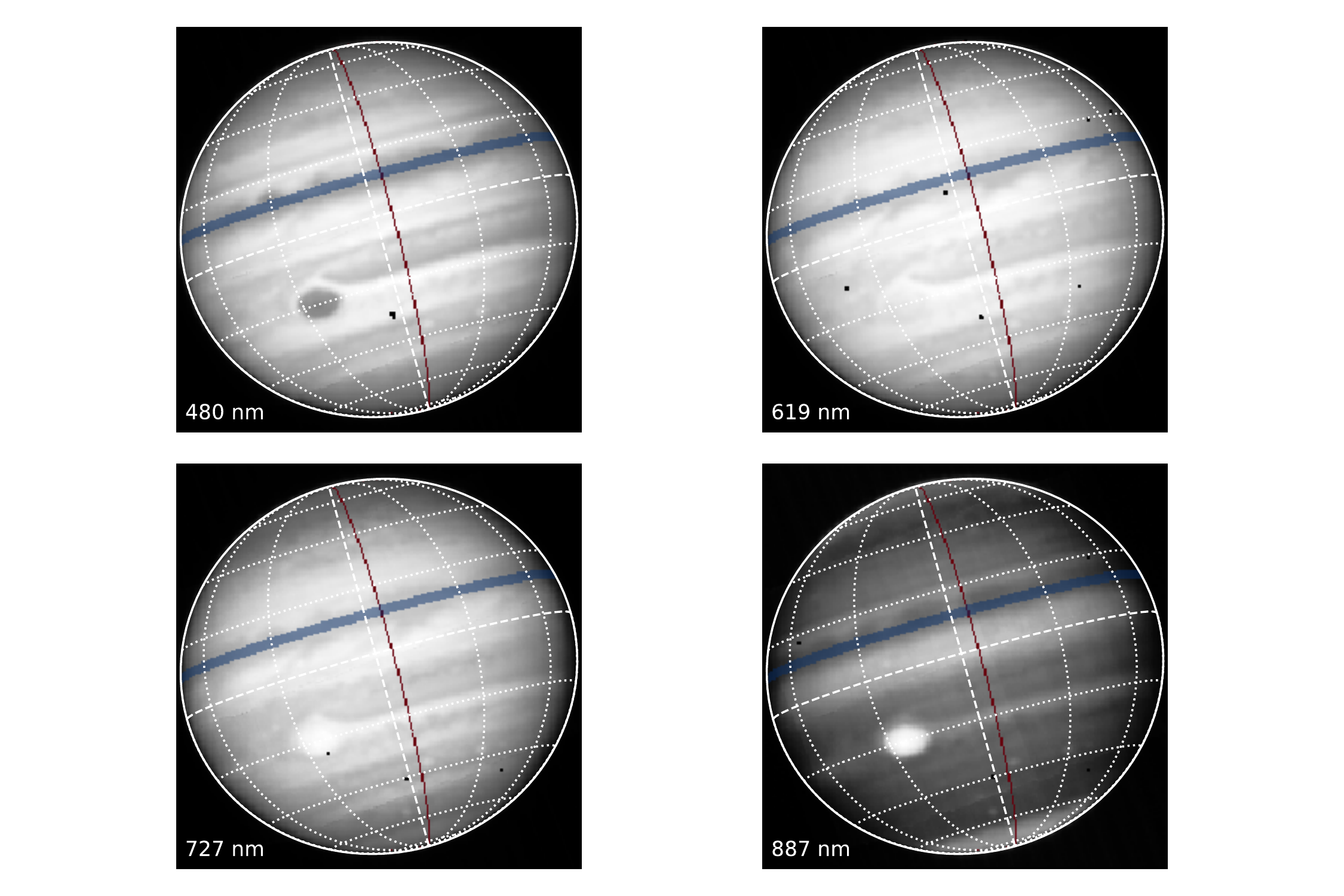}\caption{\label{fig:swath}VLT/MUSE observation of Jupiter (timestamp 2018-04-09T06:04:06.918)
shown at four individual wavelengths, with the limb-darkening swath
in section \ref{subsec:Limb-darkening-analysis} marked in blue and
the vertical swath in section \ref{subsec:Modelling-meridional-variations}
marked in red. The key to the white gridlines is as in Figure \ref{obsimages}.
We have cropped out regions of sky observed by MUSE in these images
in order to make the spatial resolution of Jupiter's surface by MUSE
clearer to the reader. Note in particular the GRS at the bottom left,
which is dark at wavelengths sensitive to chromophore but bright at
wavelengths sensitive to cloud and haze, as well as the presence of
a planetary wave in the northern NEB which is visible at all wavelengths.}
\label{limbnsswathlocation}
\end{figure}

To test the hypothesis of a universal chromophore, we first extracted
a representative latitudinal swath of the NEB between 10$^{o}$N and
13$^{o}$N (planetographic) from our primary MUSE observation in April
2018, which is shown graphically in Figure \ref{limbnsswathlocation}.
At this point in time, the southern NEB was the reddest region of
Jupiter after the GRS, was relatively homogeneous in appearance with
longitude in comparison with the South Equatorial Belt (SEB) and had
little high-altitude haze cover, making the fitting of chromophore
more straightforward to decouple from other atmospheric variables
than in other regions of Jupiter. The shape of the limb-darkening
curve along the swath was seen to be well modelled if the swath was
sampled at each wavelength at longitude values of \{$-60^{o}$, $-30^{o}$,
$0^{o}$, $30^{o}$, $60^{o}$\} relative to the central meridian,
avoiding large viewing zenith angles close to the outer perimeter
of Jupiter's planetary disc where mixing with the sky is significant.
Each longitude sample was obtained by computing a Gaussian weighted
average of all spectra within a FWHM of 3 degrees of longitude. The
respective spectral errors were then found through the standard deviation
of the Gaussian weighted average. In order to save computational time,
only 200 wavelengths per spectrum (giving 1000 wavelengths in total)
were fit in our limb-darkening analysis, selected according to both
information content \citep{Rodgers1996,ventress2014improving} and
to model the shapes of the main absorption features optimally. These
wavelengths were therefore mainly concentrated in regions of high
methane and ammonia absorption, as well as below 600 nm, while avoiding
regions of low signal-to-noise (such as around 760 nm).

\begin{figure}
\includegraphics[width=1\columnwidth,height=0.8\textheight]{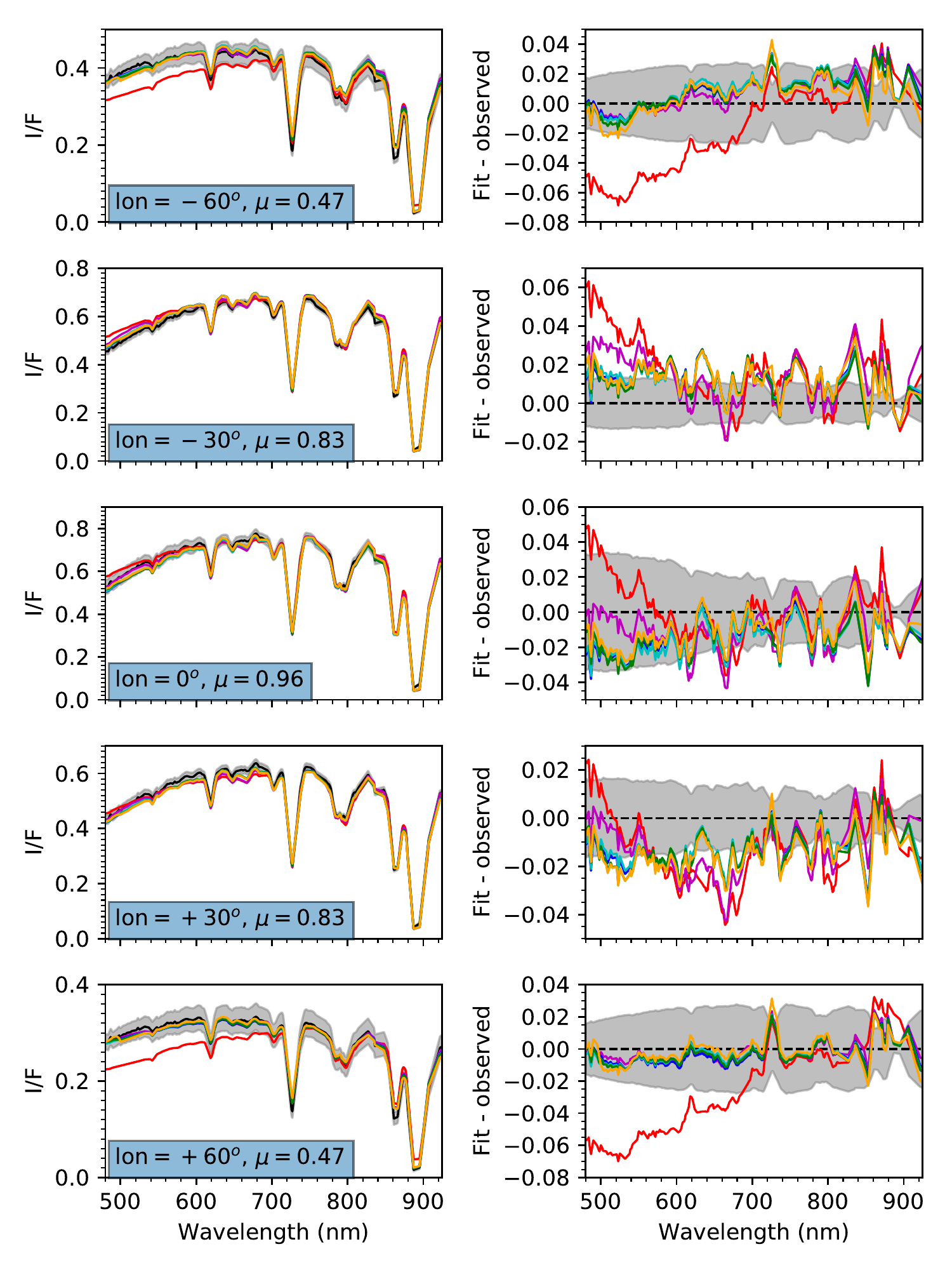}

\caption{\label{fig:limbdark}Spectral fit to the NEB at five different viewing
geometries, comparing the fit to the observed spectra (in black, with
uncertainties shaded in grey) using six different chromophore models.
In red is the fit using the \emph{Cr�me Br�l�e }model \citep{Sromovsky2017,baines2019}
with the chromophore optical constants of \citet{carlson2016chromophores}
($\chi^{2}/n=3.22$). We can see that this fails to fit the blue-absorption
gradient of the spectrum of the NEB, underestimating the slope at
the shortest wavelengths at low viewing angles while overestimating
the slope at green wavelengths at high viewing angles. By contrast,
if we use our own model, but fix $k_{c}(\lambda)$ to the optical
constants of \citet{carlson2016chromophores}, we get a much better
fit. In this particular case, shown in purple, we get the best convergence
when $n_{n}=1.42$, $r_{c}=0.1\mu m$ and $r_{n}=1\mu m$ ($\chi^{2}/n=1.48$).
Nonetheless, we can still improve on this fit. The remaining fits
use the best optical constant solutions that were retrieved directly
from the spectra, in three cases where $r_{n}=1\mu m$, $n_{n}=1.42$
and $r_{c}$ was fixed to 0.05 \textgreek{m}m (light blue, $\chi^{2}/n=1.13$),
0.2 \textgreek{m}m (dark blue, $\chi^{2}/n=1.14$), and 0.5 \textgreek{m}m
(green, $\chi^{2}/n=1.15$) respectively, plus one case where $n_{n}=1.5$,
$r_{n}=0.75\mu m$ and $r_{c}=0.5\mu m$ (orange, $\chi^{2}/n=1.17$).
All three solutions give a reasonable fit to each of the five viewing
geometries but are relatively indistinguishable in quality given the
spectral uncertainties involved.}
\end{figure}

In Figure \ref{fig:limbdark} we show clearly that the chromophore
of \citet{carlson2016chromophores} provides an inadequate fit to
the spectral slope of the NEB below 600 nm, even when accounting for
variations in particle size distribution, using the \emph{Cr\`{e}me Br\^{u}l\'{e}e
}model alone. We therefore chose to retrieve the imaginary part of
the refractive index spectrum of the chromophore, $k_{c}(\lambda)$,
directly from spectral limb-darkening of the NEB, sampled at 50 nm
intervals between 450 nm and 950 nm using the optical constants of
\citet{carlson2016chromophores} as a prior. We found that particle-size
distributions could only be weakly constrained from limb-darkening,
especially that of chromophore, but had a major effect on the retrieved
chromophore absorption spectrum. Our approach was to therefore model
all three particle populations using a gamma size distribution with
a fixed variance of 0.05, and with the effective radius of the cloud
($r_{n}$) and chromophore ($r_{c}$) particles fixed to different
values for individual retrievals chosen to maximise the parameter
space searched, and the haze effective radius ($r_{h}$) allowed to
weakly vary from a prior value of 0.5 \textgreek{m}m. These were all
selected in line with values in the literature (notably \citet{stoll1980polarimetry,ragent1998clouds,west2004jovian}
and \citet{mclean2017polarimetric}). We fixed the real parts of the
refractive index spectra of both the chromophore ($n_{c}$) and haze
($n_{h}$) particles to 1.4 at a wavelength of 700 nm, but fixed that
of the cloud particles to several different values from 1.42 (equivalent
to ammonia ice according to \citet{martonchik1984optical}) to 1.6.%
{} A detailed key of the fixed and variable parameters of our model
is shown in Table \ref{tab:Explanation-of-variables}.

\begin{figure}
\includegraphics[width=1\textwidth,height=0.9\textheight,keepaspectratio]{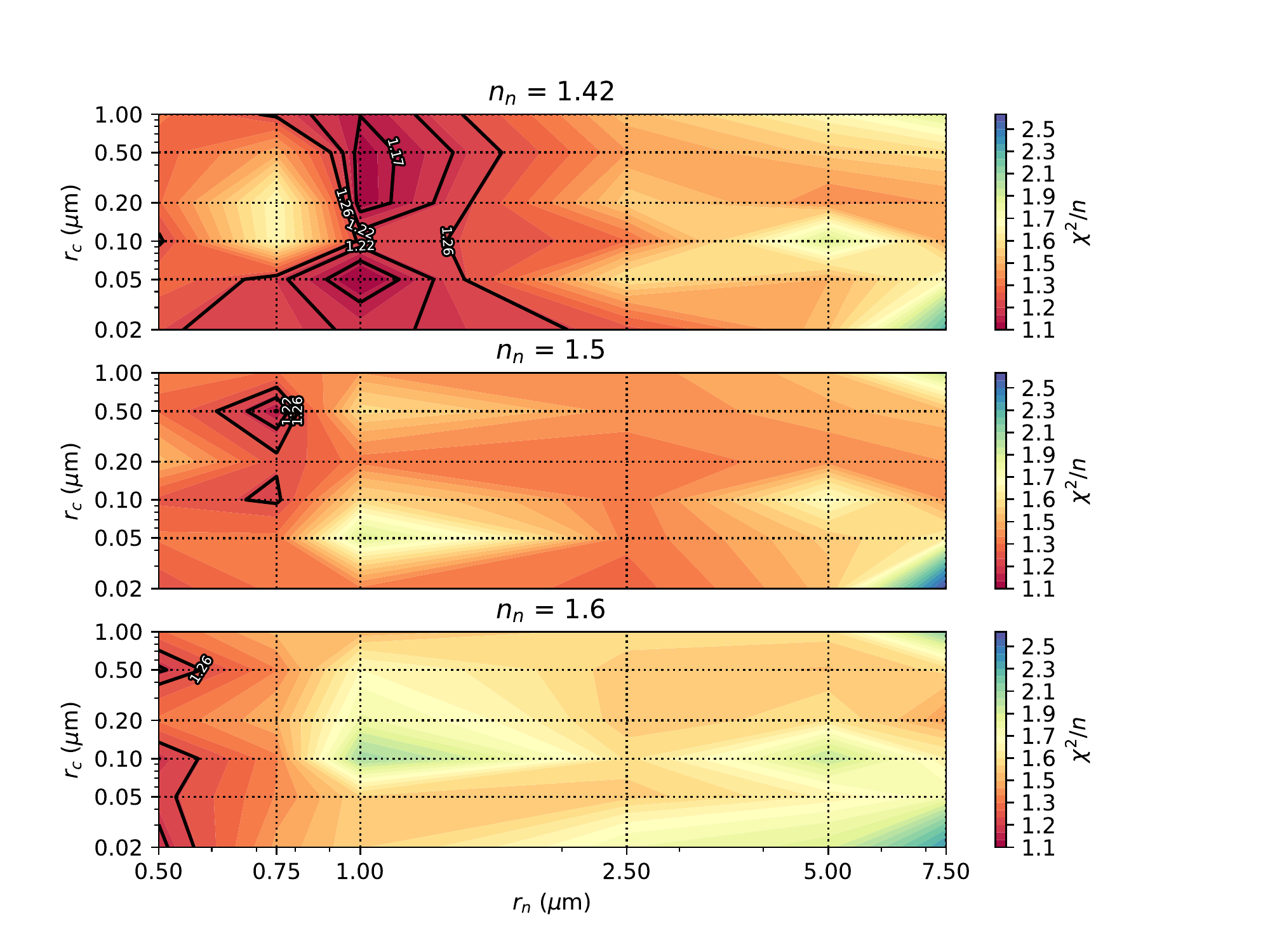}

\caption{\label{fig:chisquared}A map of the `goodness of fit' ($\chi^{2}/n$)
values obtained through limb-darkening analysis of the NEB, as a function
of the effective radius of cloud ($r_{n})$ and chromophore ($r_{c}$)
particles, as well as the real part of the refractive index $n_{n}$
of the cloud particles. Individual values of prior $r_{n}$ and $r_{c}$
used in our retrievals are marked at the intersections of each of
the vertical and horizontal black lines, and the resulting contours
interpolated from those points. Dark red colours indicate a better
quality of fit in each case. The lowest $\chi^{2}/n$ value we were
able to retrieve was equal to 1.12, and the solid black lines give
the $\chi^{2}/n$ values within a 1-, 2- and 3-sigma uncertainty interval
respectively around this minimum value. In general, for each value
of $n_{n}$, the optimal solutions tend to cluster at a given value
of $r_{n}$ but for which a wide range of $r_{c}$ values are possible.
Raising the value of $n_{n}$ tends to lower this optimal value of
$r_{n}$. However, two local minima in $\chi^{2}/n$ are clearly present
in the $n_{n}=1.42$ case that are not as profound for higher values
of $n_{n}$. This provides some evidence that the deepest cloud layers
observable at visible and near-infrared wavelengths have a low real
refractive index value.}
\end{figure}

We were unable to fit the swath with a goodness of fit value ($\chi^{2}/n$,
where n = 1000) less than 1 for any prior parameter, and within the
parameter space searched in this analysis, only solutions where $r_{n}=1.0\mu m$
and $n_{n}=1.42$ could provide a fit with $\chi^{2}/n$ < 1.2. Of
these, two sets of solutions were found to be applicable to within
uncertainty of the lowest $\chi^{2}/n$ value (where the uncertainty
is given by $\sigma_{\chi^{2}/n}=\sqrt{2/n}=0.45$): one where $r_{c}<0.1\mu m$
and one where $r_{c}>0.5\mu m$, as shown in Figure \ref{fig:chisquared}.
Raising $n_{n}$ a small amount could also provide reasonable $\chi^{2}/n$
values so long as $r_{n}$ was lowered to compensate, but if $n_{n}$
was set to 1.6 or above we obtained very poor fits at high viewing
zenith angles to the methane absorption feature at 619 nm regardless
of particle size distribution. This was contrary to the findings of
\citet{sato2013retrieval} who obtained a value of $n_{n}$ closer
to 1.85 in the South Tropical Zone (STropZ) from Cassini/ISS measurements.
We should note that \citet{Howett2007} state a real refractive index
for NH\textsubscript{4}SH of 1.75 at visible and near-infrared wavelengths,
but they only provide citations of this value through personal communication.
Nonetheless, if this value can be independently verified, then it
would indicate that the deepest cloud layer we observe in the visible
and near-infrared is not made primarily of NH\textsubscript{4}SH,
at least in the NEB. %

\begin{figure}
\includegraphics[width=0.7\textwidth]{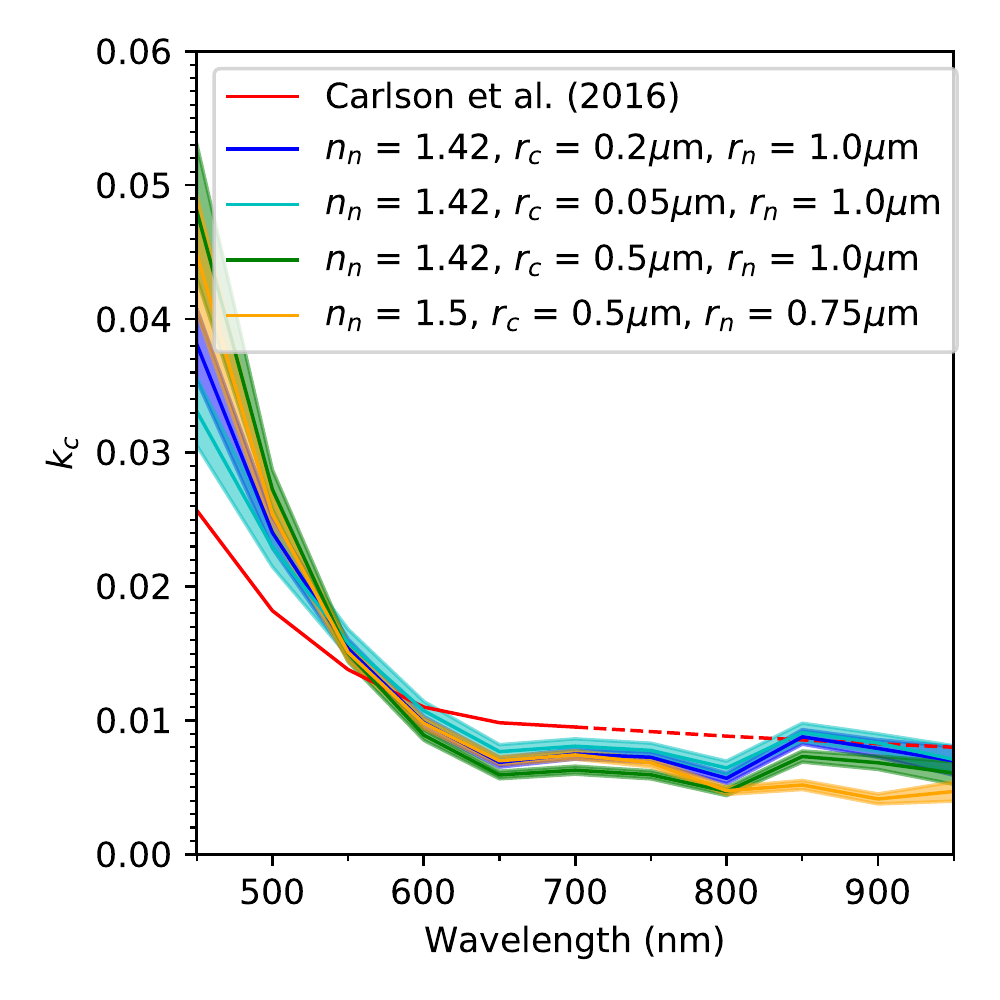}

\caption{Chromophore imaginary refractive index solutions $k_{c}(\lambda)$
retrieved through limb-darkening analysis of the NEB (showing the
four solutions with the best $\chi^{2}/n$ values) compared with the
tabulated optical constants of \citet{carlson2016chromophores}, with
the legend as in Figure \ref{fig:limbdark} with uncertainties shaded.
We can see that the general shape of the retrieved solutions of $k_{c}(\lambda)$
is relatively invariant regardless of $r_{c}$, with only minor modifications
needed to the spectral slope.}
\label{fig:imagri}
\end{figure}

Although the retrieved chromophore imaginary refractive index spectrum
changes substantially with $r_{c}$, our solutions all have several
features in common. In all cases, we require a spectral slope between
476 nm and 600-650 nm that is greater than can be provided by the
chromophore of \citet{carlson2016chromophores}, as shown in Figure
\ref{fig:imagri} and Table \ref{imagritable}. Usually, our model
does this by raising the imaginary refractive index at the shortest
wavelengths, and decreasing it at longer wavelengths relative to \citet{carlson2016chromophores}.
We also retrieved a secondary absorption peak around 850 nm, although
this could be a result of uncertainties in methane band absorption
data as opposed to a genuine chromophore absorption feature. However,
we should clarify that the experimental uncertainties on the optical
constants of the chromophore of \citet{carlson2016chromophores} were
not published, and so it is possible that the uncertainty in the thickness
of the chromophore film produced in the laboratory, or some other
factor, may affect the value of the imaginary refractive index spectrum
they derive. %
We therefore cannot rule out the possibility of red colour on Jupiter
forming in a manner similar to that of the chromophore of \citet{carlson2016chromophores}.

We should note that, using the model described in section \ref{subsec:Cloud-and-chromophore},
but still fixing $k_{c}(\lambda)$ to the optical constants of \citet{carlson2016chromophores},
we were still able to provide a much better fit to the spectra than
using the \emph{Cr\`{e}me Br\^{u}l\'{e}e} model, with the best results obtained
when $r_{n}=1\mu m$, $r_{c}=0.1\mu m$ and $n_{n}=1.42$, and with
a fit to the spectrum not far removed from the case where we retrieved
$k_{c}(\lambda)$ directly, with $\chi^{2}/n=1.48$. Nonetheless,
we chose to stick with our own retrieved chromophore absorption spectrum
for a number of reasons. Firstly, our retrieved chromophore absorption
spectrum did provide some improvement to the fit to the shape of the
blue-absorption gradient, particularly at lower viewing zenith angles.
We also remark on the fact that the increase in the retrieved blue-absorption
gradient relative to the chromophore of \citet{carlson2016chromophores}
is consistent regardless of the values of $r_{n}$, $r_{c}$ or $n_{n}$
we choose. However, we find that the greatest justification for retrieving
the chromophore absorption spectrum directly is through fitting of
the GRS, as we describe in the next section.

\subsection{Selection of a universal chromophore\label{subsec:Selection-of-universal}}

In order to confirm whether any chromophore solution retrieved from
the NEB could be deemed that of a `universal chromophore', we must
by definition be able to apply the same chromophore solution to any
spectrum of Jupiter and obtain a fit to the spectral slope below 600
nm that is adequate. In addition, the resulting retrieved aerosol
abundances should be in line with theoretical expectations as well
as with other prior observations of Jupiter. We first attempted to
fit each `well'-retrieved chromophore imaginary refractive index solution,
defined as those for which $\chi^{2}/n<1.3$ in the NEB and for which
$r_{h}<r_{n}$ (in order to be physical), to a single spectrum extracted
from the centre of the GRS as observed in April 2018. This spectrum
was selected as it had the greatest spectral slope at blue wavelengths
found anywhere on Jupiter, and was the most difficult spectrum to
model using the chromophore of \citet{carlson2016chromophores}. Unlike
with the aforementioned limb-darkening analysis, we fit all 435 wavelengths
between 476 nm and 910 nm of the spectrum of the GRS; wavelengths
above 910 nm were seen to be overly affected by second-order contamination
and were therefore cropped out. As in the previous section, a detailed
list of fixed and variable parameters is shown in Table \ref{tab:Explanation-of-variables}.
In order to better compare our retrieved aerosol density values with
those in the literature, usually quoted in units of g/l, we have assumed
cloud, haze and chromophore particle mass densities to all be equal
to 0.87g/cm\textsuperscript{3}, equivalent to that of ammonia ice
\citep{satorre2013refractive}. A corresponding mass density value
for N\textgreek{H}\textsubscript{4}SH is lacking in the literature,
but we assume that it is approximately triple that of NH\textsubscript{3}
ice based on the difference in molecular weight.

As shown in Table \ref{tab:grssolutions}, we found that it was often
simpler to fit chromophore solutions with higher values of $n_{n}$
to the GRS than it was with lower values. However, we find that the
retrieved cloud densities also drop dramatically with increasing $n_{n}$:
as one increases the value of $n_{n}$ from 1.4 to 1.6, the retrieved
cloud densities drop by approximately a factor of 10 for given values
of $r_{n}$ and $r_{c}$, and in practice, given the lower best-fit
values of $r_{n}$ for solutions with increasing $n_{n}$, the decrease
in retrieved cloud density is even more substantial. \citet{Palotai2014}
estimated consistent aerosol mass densities in the centre of the GRS
of the order of $10^{-5}$ g/l over all visible altitudes using a
general circulation model, while \citet{zuchowski2009} estimated
average NH\textsubscript{4}SH cloud mass densities in the STropZ,
one of the cloudiest bands on Jupiter, of the order of $10^{-6}$
g/l. By contrast, the Galileo Probe Nephelometer retrieved maximum
mass densities of the order of $10^{-7}-10^{-6}$ g/l \citep{ragent1998clouds}
at its entry site in a hotspot region where aerosol densities are
predicted to be exceptionally low by Jovian standards, and certainly
lower than the GRS. This makes it very difficult to justify the retrieved
values of cloud density of the order of $10^{-8}$-$10^{-7}$g/l in
the GRS if $n_{n}$ is increased well above that predicted for ammonia
ice, or even to 1.5 which is within the margin of error of laboratory
measurements of the refractive index of ammonia ice (quoted as 1.48
in both \citet{Romanescu2010} and \citet{satorre2013refractive}).
Prior constraints from the literature, together with our own from
limb-darkening in the NEB, would therefore rule out a high $n_{n}$
solution even if they provided a good fit to the GRS (we did not test
solutions for $n_{n}$ lower than 1.42 due to the absence of substances
likely to be present in Jovian clouds with such a low refractive index,
other than water ice which freezes out of the atmosphere at pressure
levels not observable in the MUSE spectral range). We also note that
the higher the value of $r_{c}$, the greater the retrieved mass density
of the cloud layer. However, we obtained an inferior fit to the spectrum
of the GRS if we raised $r_{c}$ to 0.5 \textgreek{m}m. As a compromise,
we therefore chose our `universal chromophore' solution as the one
where $r_{n}=1\mu m$, $r_{c}=0.05\mu m$ and $n_{n}=1.42$. This
resulted in a cloud mass density of approximately $(1.7\pm0.1)\times10^{-6}$
g/l - on the lower end of predicted mass densities for Jovian clouds,
but not entirely unrealistic.

\begin{figure}
\includegraphics[width=1\columnwidth]{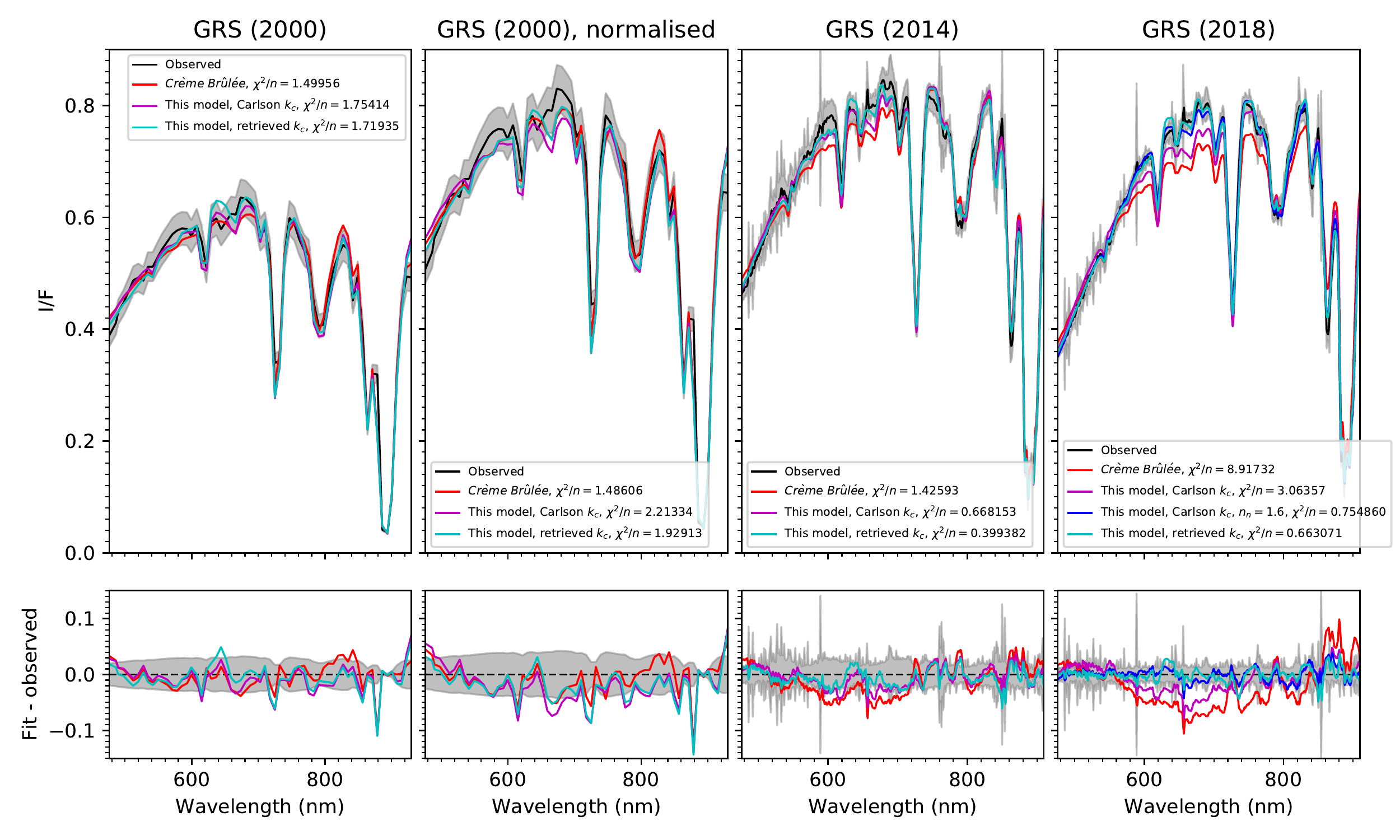}\caption{\label{fig:grsspecfits} A comparison of the fit to the spectrum of
the centre of the GRS obtained at different stages in its evolution:
(left) in 2000, as obtained by Cassini/VIMS and provided in the supplementary
data of \citet{carlson2016chromophores}; (second from left) the same
spectrum, but normalised to an I/F value of 0.83 at 673 nm as suggested
by \citet{carlson2016chromophores} in order to be in line with contemporaneous
observations of the GRS (although \citet{Sromovsky2017} suggest an
intermediate scaling factor of 1.12 times the spectrum on the far
left, based on comparisons of disc-averaged spectra of Jupiter); (second
from right) in 2014, as obtained by VLT/MUSE before the simultaneous
shrinking and reddening events described in \citet{simon2014dramatic,Simon2018};
and (right) in 2018, as obtained by VLT/MUSE after the shrinking and
reddening of the GRS. In the case of the GRS (2000) spectra, we smoothed
the methane k-tables and solar spectra to 7 nm resolution to be in
keeping the VIMS-V spectral sampling before performing our retrievals.
In each of the four cases, we compare the fits using the \emph{Cr�me
Br�l�e} model in red, our own model but fixed to the optical constants
of \citet{carlson2016chromophores} in purple, and using our own model
but with our own directly-retrieved optical constants in light blue
(using the highlighted `best-fit' values of $n_{n}$ and $r_{c}$
in Table \ref{retrievalresults}). Although our retrieved optical
constants provide a superior fit in all three cases, the fit using
the \emph{Cr�me Br�l�e }model is still passable in 2000 and 2014 albeit
with some discrepancy between 600 nm and 700 nm. This may explain
the reasonable fits found by \citet{Sromovsky2017} and \citet{baines2019}
to VIMS spectra of the GRS using the\emph{ Cr�me Br�l�e }model. By
2018, however, it is impossible to fit the spectrum of the GRS using
the optical constants of \citet{carlson2016chromophores}, unless
$r_{n}$ is raised to 1.6, which we show in dark blue ($r_{c}=0.05\mu m$,
$r_{n}=5\mu m$). This clearly explains our motivation to retrieve
our own chromophore solution.}
\end{figure}

The fit of this optimal chromophore solution to spectra of the GRS
before and after the reddening events reported by \citet{simon2014dramatic,Simon2018}
are shown in Figure \ref{fig:grsspecfits}. We found the fits to be
adequate in both cases, unlike those provided by the optical constants
of \citet{carlson2016chromophores} particularly following the most
recent reddening event. The only way in which we are able to fit the
GRS spectrum in 2018 using the optical constants of \citet{carlson2016chromophores}
is to raise $r_{n}$ to 1.6, but this would result in the underestimation
of cloud opacity for reasons previously discussed. This therefore
begs the question as to why our results contradict those of \citet{Sromovsky2017}
so substantially, who were able to provide a good fit to their own
Cassini/VIMS spectra of the GRS using the chromophore of \citet{carlson2016chromophores}.
The most likely reason is that, while the most dramatic reddening
has only occurred in the past few years, the blue-absorption gradient
of the spectrum of the GRS had been gradually but consistently steepening
at least since the Voyager era. This means that VIMS spectra of the
GRS obtained during the Cassini flyby in 2000, as well as MUSE spectra
obtained in 2014, were substantially easier to fit using the \citet{carlson2016chromophores}
optical constants than MUSE spectra from 2018, as we also show clearly
in figure \ref{fig:grsspecfits}. Nonetheless, we should also note
that the general fit to the Cassini/VIMS spectrum of the GRS from
2000, as calibrated and supplied by \citet{carlson2016chromophores},
is poorer than to the MUSE spectra from 2014 and 2018, regardless
of the model used. We believe that this can mostly be blamed on deficiencies
in the VIMS calibration. The shape of the methane absorption features
at 727 nm and 880 nm in particular are both unphysical and difficult
to fit, even when taking into account the lower spectral resolution
relative to MUSE in the reference gas absorption data, and this is
clearly seen when compared to the smoothed disc-averaged spectra of
\citet{karkoschka1994spectrophotometry,karkoschka1998methane} by
both \citet{Sromovsky2017} and \citet{baines2019}.

In all cases, we retrieved relatively consistent chromophore abundances
in the GRS regardless of $r_{c}$ of the order $10^{-8}$ g/l, which
is also in agreement with the findings of \citet{Sromovsky2017},
but would require a flux rate of acetylene into the tropopause to
produce that is at least a factor of 10 higher than that predicted
through photochemistry alone \citep{Moses2010}.

\section{Application of universal chromophore}

\subsection{Fits to representative spectra}

\begin{figure}
\includegraphics[width=1\columnwidth,height=0.8\textheight,keepaspectratio]{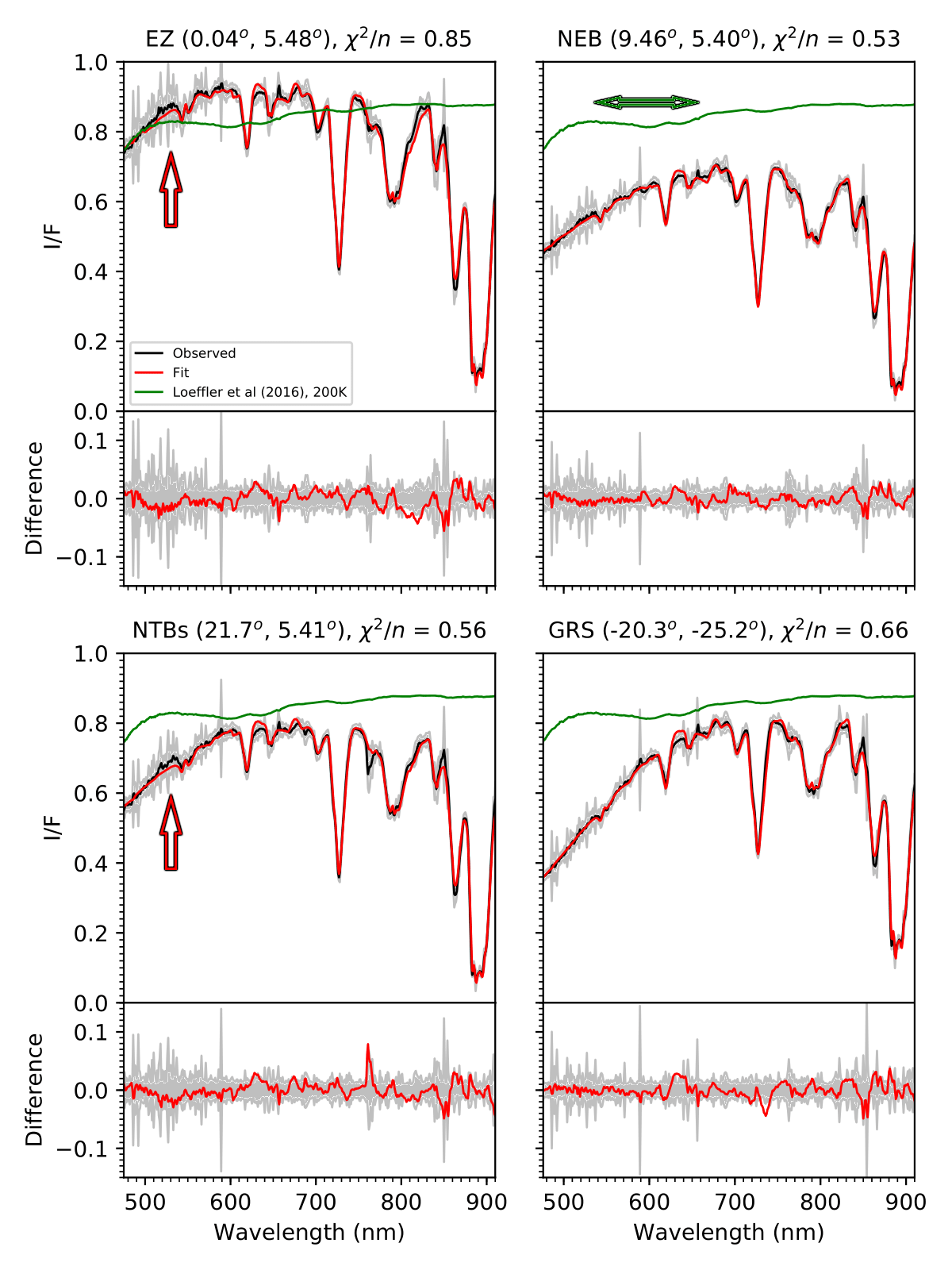}

\caption{\label{fig:fitcomparison}Fits to different representative MUSE spectra,
using the `universal chromophore' optical constants selected in section
\ref{subsec:Selection-of-universal}. The NTBs spectrum was obtained
from MUSE dataset 2017-05-15T02:01:59.328 when the red colour of the
NTBs was at its most prominent, while the remaining three spectra
were obtained from dataset 2018-04-09T06:04:06.918 which was taken
under the best observing conditions of all three datasets. For each
of the four spectra, the given latitude values are planetocentric,
while longitude values are east relative to the sub-observer. $\chi^{2}/n$
values of the quality of the fit of the retrieved imaginary refractive
index spectrum to each MUSE spectrum are also shown. We have highlighted
small deviations in the fit to the EZ and NTBs spectra at blue wavelengths
with the red arrows, for which we have recommended further investigation
(the apparent absorption feature present around 760 nm is due to telluric
noise, which is particularly prominent in the NTBs spectrum as a consequence
of imperfect airmass correction). For comparison, we also show the
shape of the NH\protect\textsubscript{4}SH absorption spectrum (in
green, with the extent of the 600 nm absorption feature highlighted
by the green arrow in the top-right diagram) as obtained by \citet{Loeffler2016}
when reheated to 200K following proton irradiation, which clearly
models the shape of Jovian spectra very poorly. Reference NH\protect\textsubscript{4}SH
absorption spectra obtained at lower temperatures have a similar shape
to that at 200K, but with a much deeper 600 nm absorption feature
that is non-existent in Jovian spectra.}
\label{specfits}
\end{figure}

\begin{figure}
\includegraphics[width=1\textwidth,height=1\textheight,keepaspectratio]{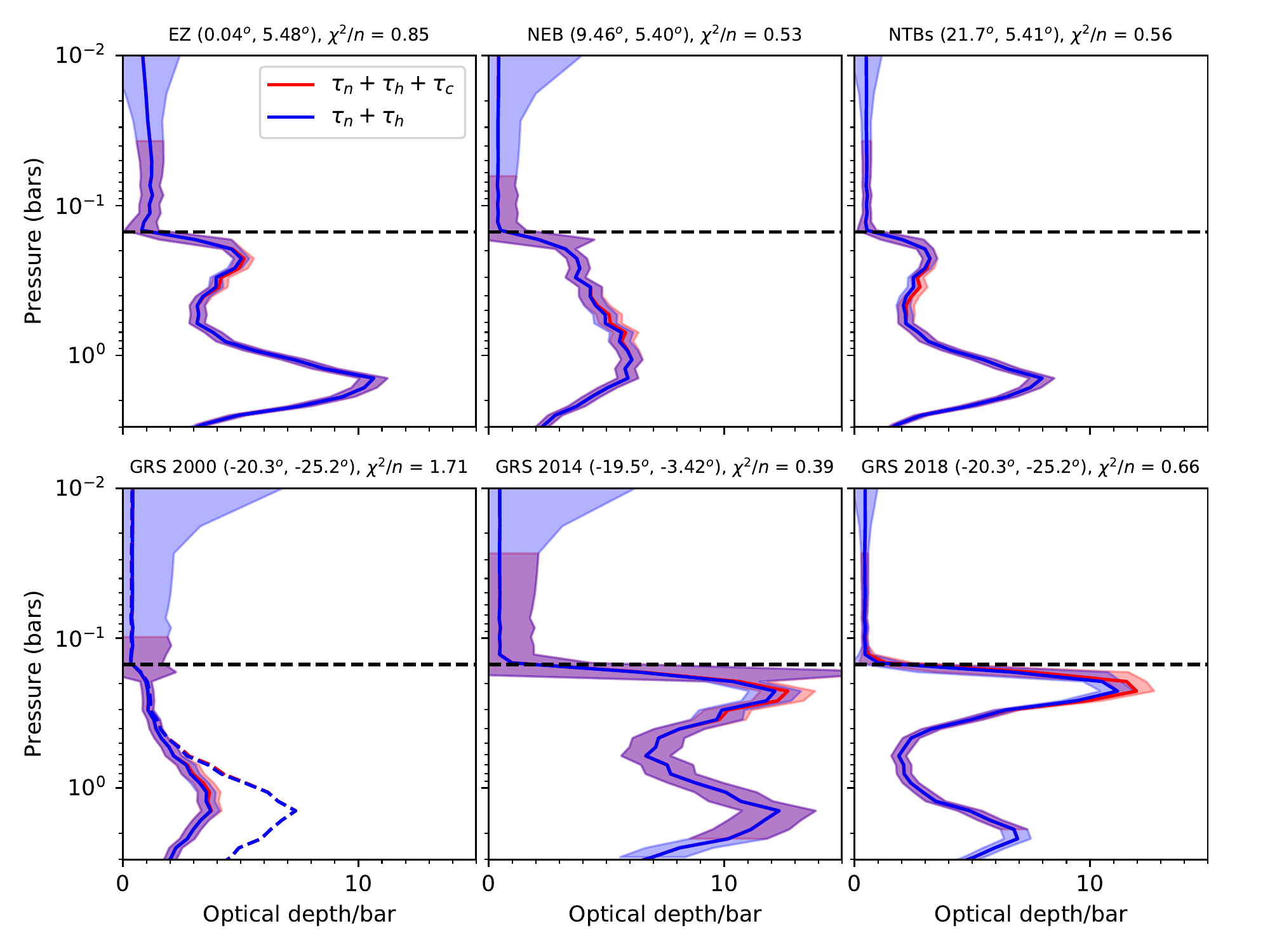}

\caption{Retrieved cloud, chromophore and haze vertical profiles for the EZ,
the NEB, the NTBs (in 2017) and the GRS from three different years,
with errors shaded. The GRS (2000) solid profiles were retrieved using
the unnormalised VIMS spectrum, with the dashed lines showing the
results for the normalised spectrum. In each plot, the black dashed
line shows the 0.15 bar cut-off between the $\tau_{n}$ and $\tau_{h}$
profiles.}

\label{cloudprofiles}
\end{figure}

We now discuss the fit of our universal chromophore solution in relation
to spectra from three different locations on Jupiter other than the
GRS, as shown in Figure \ref{fig:fitcomparison}, and with retrieved
cloud and chromophore vertical profiles given in Figure \ref{cloudprofiles}.
Each was chosen to represent the three main spectral morphologies
one typically finds on Jupiter (excluding polar and hotspot regions):
high continuum reflectivity with low blue-absorption (the Equatorial
Zone, or EZ for short), low continuum reflectivity with high blue-absorption
(the NEB), and high continuum reflectivity with high blue-absorption
(the NTBs and GRS). Wavelengths below 600 nm were fit well within
the uncertainty boundaries for the belts and the GRS. When it comes
to the EZ, however, we can see that the fit to shorter wavelengths
is noticeably inferior to other regions of Jupiter, even when variations
in particle size are accounted for. We focus on the spectral region
that consistently caused the greatest perturbation to the blue-absorption
slope, which is between approximately 510 and 540 nm, where we discern
a small increase in I/F in the EZ and the NTBs relative to the NEB
and the GRS, which was not accounted for in our retrieved chromophore
absorption spectrum. Unfortunately, this is a wavelength region with
a somewhat low signal-to-noise ratio (averaging around 20-30 between
510 and 540 nm, as opposed to \textasciitilde 60-100 at continuum
near-infrared wavelengths) due to the uncertainty in the interpolation
of stellar lines when calibrating the spectra. Nonetheless, we assume
this variation is genuine as it does not appear to be either random
or systematic over the surface of Jupiter. The variability in I/F
at these wavelengths could be due to a number of factors: a) a genuine
secondary chromophore absorption feature (perhaps due to an $n\rightarrow\pi^{*}$
transition \citep{Nassau1983}) that could be characteristic of a
certain molecular endmember, b) a consequence of other properties
of the atmosphere that we have not properly accounted for, but which
might alter the local blue-absorption gradient (such as the presence
of additional aerosol or gas layers, or local changes in temperature
and pressure that could alter the molecular structure of the chromophore
and hence the breadth of its primary absorption feature), or c) poor
band data around the 540 nm methane absorption feature. This particular
feature aside, however, our spectral fits show that we cannot entirely
rule out the possibility of a universal chromophore. To either confirm
or reject this universal chromophore solution definitively, we would
require spectral information at wavelengths shorter than the lower
limit of 475 nm provided by MUSE's nominal wavelength range.

For comparison, we have also included absorption spectra of irradiated
NH\textsubscript{4}SH, itself a candidate chromophore as proposed
by \citet{Lebofsky1976} and more recently by \citet{loeffler2015giant,Loeffler2016}.
We can see that the shape of these absorption spectra model blue-absorption
poorly on Jupiter for two main reasons. One is that the reflectance
spectrum of irradiated NH\textsubscript{4}SH only appears to show
substantial absorption below 500 nm, whereas the shortwave absorption
peak in Jovian spectra extends all the way to 600 nm. This discrepancy
may be alleviated by altering the altitude of the irradiated NH\textsubscript{4}SH
particles or changing their size distribution, but this is difficult
to verify due to the absence of NH\textsubscript{4}SH optical constant
data below 1000 nm in the literature. The other issue pertains to
the presence of a secondary broad absorption peak at around 610 nm
due to $S_{3}^{-}$ radicals, which is completely absent in Jovian
spectra. This extra absorption peak is known to disappear under only
two circumstances: a) the NH\textsubscript{4}SH is irradiated at
temperatures far below even the coldest temperatures found on Jupiter
\citep{Loeffler2018}, or b) the NH\textsubscript{4}SH is irradiated
at temperatures typically found in the upper troposphere, but is then
reheated to temperatures one would only expect around 2 bars, substantially
below the maximum visible penetration depth even in the belts \citep{Loeffler2016}.
We will therefore rule out irradiated NH\textsubscript{4}SH as a
candidate chromophore in this work until a plausible hypothesis can
be found to explain these two discrepancies.

\subsection{Modelling meridional variations in tropospheric aerosol structure\label{subsec:Modelling-meridional-variations}}

\begin{figure}
\includegraphics[width=1\textwidth,height=0.9\textheight,keepaspectratio]{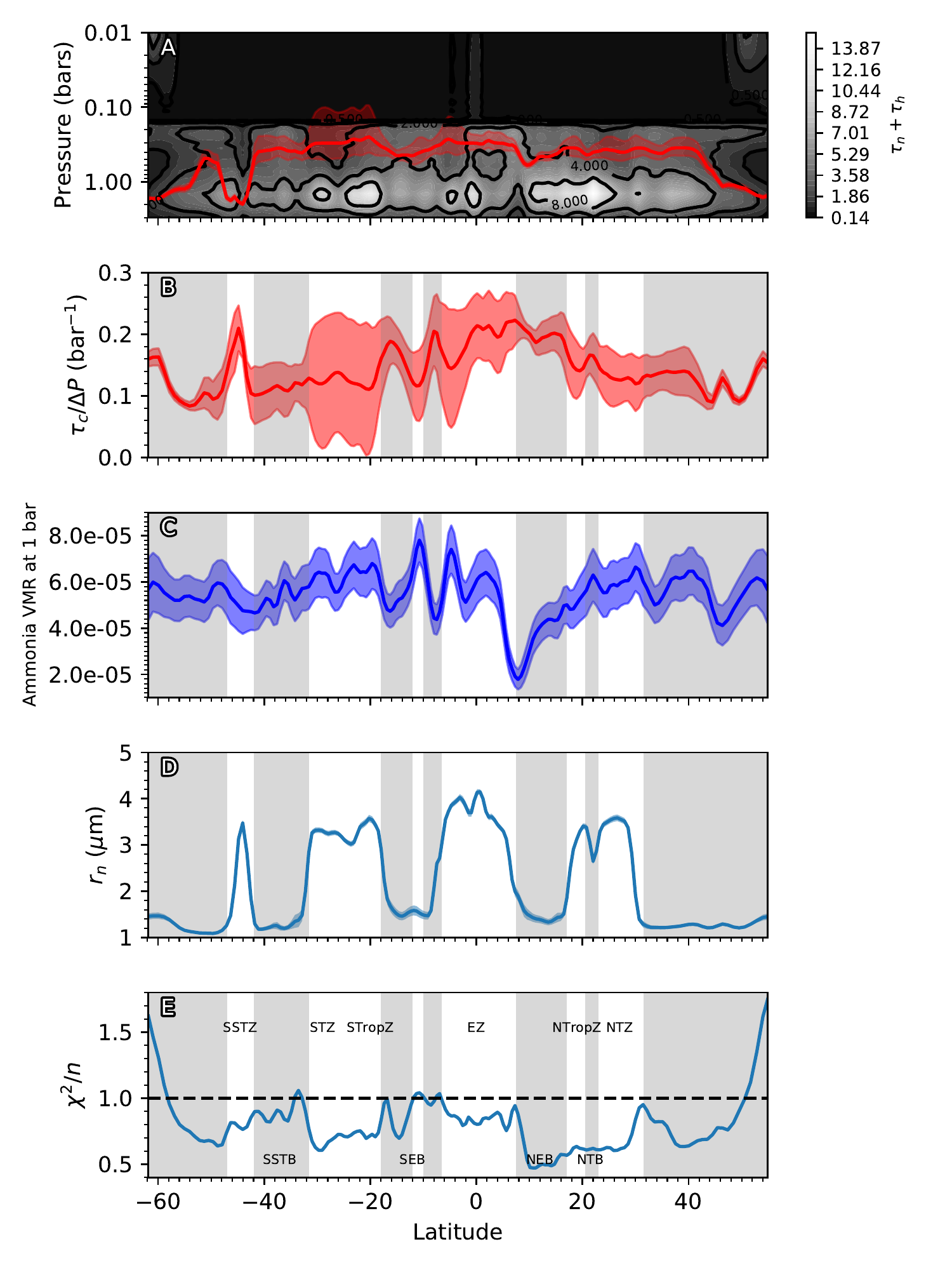}

\caption{\label{fig:nsswath}Results of the retrieval of variations in cloud,
haze, colour and ammonia gas over the swath of Jupiter shown in Figure
\ref{fig:swath}, as a function of planetocentric latitude. Gaussian
smoothing over latitude was performed for these plots in order to
remove pixel-to-pixel fluctuations in the retrieved parameters and
instead observe general trends. Plot A shows the vertical profile
of cloud ($\tau_{n}/\Delta p$, $P>0.15$ bars) and haze ($\tau_{h}/\Delta p$,
$P<0.15$ bars) abundance in units of optical depth/bar, with iso-contours
at 0.5, 1, 2, 4 and 8 bar\protect\textsuperscript{-1} respectively.
In red is shown the retrieved chromophore altitude $P_{c}$. Plot
B shows the retrieved chromophore abundance $\tau_{c}/\Delta p$,
also in units of optical depth/bar, with errors shaded in light red.
Plot C shows the retrieved gaseous ammonia profile sampled at 1 bar,
approximately the level of greatest sensitivity to ammonia gas based
on the retrieved pressure levels of the cloud layers, with uncertainties
shaded in light blue. Plot D shows the retrieved cloud particle effective
radius $r_{n}$ in \textgreek{m}m. Finally, Plot E shows the goodness
of fit ($\chi^{2}/n$) to the spectrum at each latitude. The approximate
boundaries between the zones (in white) and belts (in grey) are marked
in plots B-E, with the names of the respective zones and belts marked
in plot E.}
\end{figure}

Having established that our universal chromophore solution could provide
a mostly good fit to representative spectra of Jupiter, we then applied
this solution more generally to look at overall meridional variations
in cloud structure, chromophore abundances and cloud-top gaseous ammonia
abundances on Jupiter. We extracted a swath of single-pixel width,
as shown at different wavelengths in Figure \ref{fig:swath}, at a
longitude chosen to minimise the viewing zenith angle of observation
of each latitude, but also to prevent the bisection of longitudinally
anomalous regions in order simply to observe `generic' zone-belt variations
(although in practice, given the turbulent nature of the northern
NEB at this time, this was not entirely possible). We did not analyse
latitudes for which the viewing zenith angle was greater than $60^{o}$,
for several reasons: a) to save computational time (due to the increased
number of quadrature points required to perform integration over viewing
zenith and azimuthal angles closer to the limb), b) to avoid spectral
mixing with the sky, a consequence of MUSE's relatively low spatial
resolution and uncertainties in the PSF, and c) we were unable to
model variations in the spectral fit around 890 nm in the polar regions
(as can be seen in Figure \ref{fig:nsswath}), even if we accounted
for a thick stratospheric `polar hood'. In total, we therefore performed
spectral analysis on 173 individual spectra in the swath, obtained
at single-pixel resolution. We performed this analysis only on a single
Jupiter observation (2018-04-09T06:04:06.918) due to its optimal observing
conditions and calibration, even though this observation took place
during notable meteorological events for which retrieved atmospheric
parameters in certain latitudinal bands (particularly the EZ and NEB)
may have been unique to that particular time. A broader analysis of
temporal changes in colour and cloud structure between 2014 and 2018
is beyond the scope of this paper, as are variations in cloud-top
ammonia gas abundances which are highly model-dependent and variable
in time and longitude. We refer the reader instead to the thesis of
\citet{mythesis} for more information on these changes.

\begin{figure}
\includegraphics[width=1\textwidth]{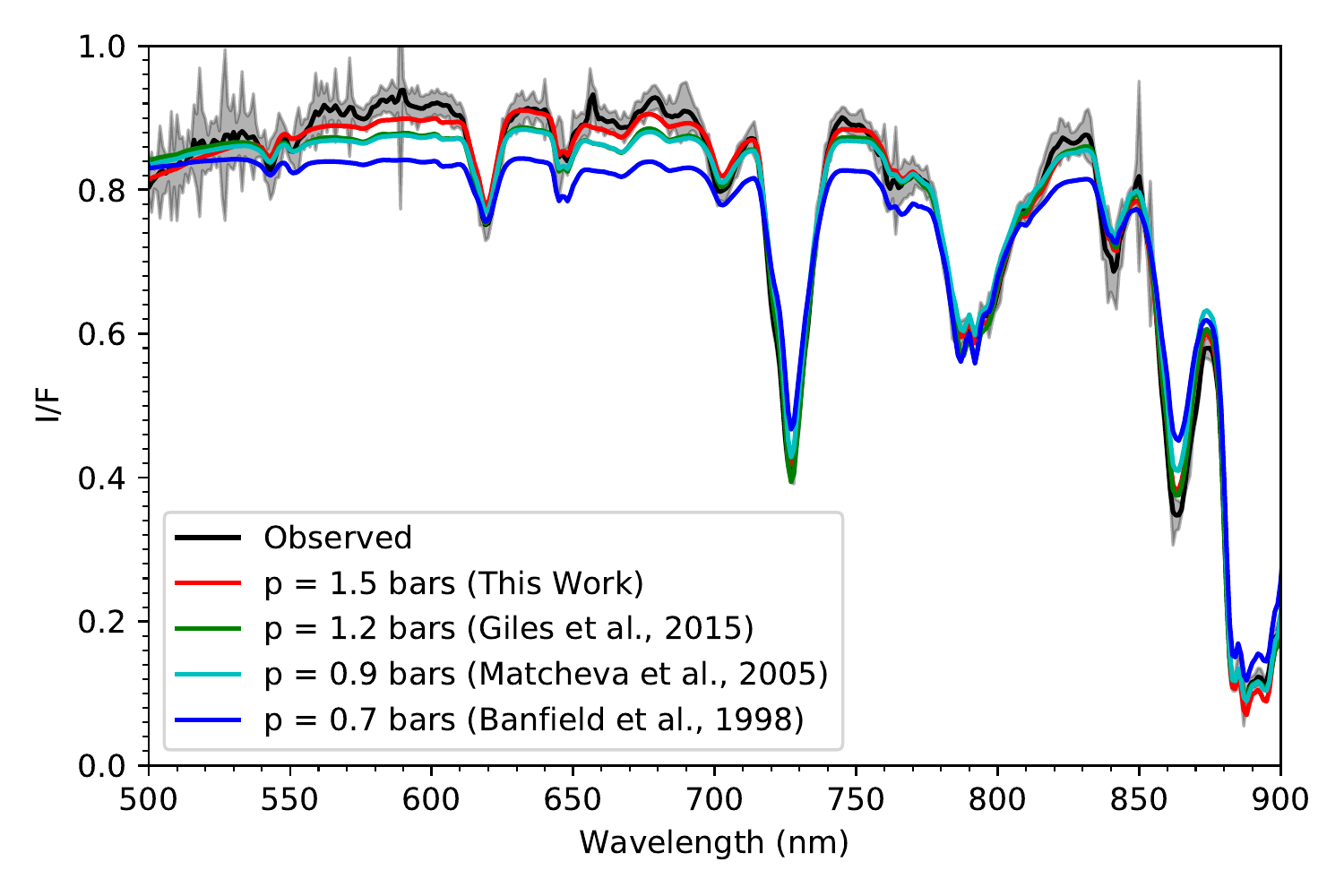}

\caption{\label{fig:ezcloudlevel}Spectral fit to the EZ (lat, lon = 2.86$^{o}$,
5.16$^{o}$) when $\tau_{n}$ is set to fall to 0 deeper than four
different pressure levels as stated in the legend, in order to simulate
the base level of the deepest visible cloud layer as found by \citet{banfield1998jupiter}
from Galileo/SSI data in the near-IR, \citet{matcheva2005cloud} from
Cassini/CIRS observations in the mid-IR, and \citet{Giles2015} from
Cassini/VIMS-IR data at 5 \textgreek{m}m. We find that the best fit
to the EZ is provided when the deepest visible cloud layer is located
around 1.5 bars ($\chi^{2}/n=1.31$), while placing the cloud base
at 1.2 bars ($\chi^{2}/n=2.67$), 0.9 bars ($\chi^{2}/n=3.94$) or
especially at 0.7 bars ($\chi^{2}/n=11.1$) underestimates I/F at
visible continuum wavelengths relative to wavelengths of major methane
absorption. The reader should note that the \citet{Giles2015} and
\citet{matcheva2005cloud} retrievals match very closely, except at
the 860 nm methane absorption feature.}
\end{figure}

The results of our analysis are shown in Figure \ref{fig:nsswath}.
As the belts are thought to be regions of net downwelling, we would
expect the altitude of the cloud layer to be deeper here than in the
zones. In addition, a lack of cloud and haze cover in the belts would
allow for the observation of deeper layers than in the zones. However,
we observe a relative lack of variation in the altitude of the main
cloud layer, usually retrieved around (1.4$\pm0.3)$ bars in both
the zones and the belts (although the altitude of the cloud in the
belts is more poorly-defined, as can be seen in Figure \ref{cloudprofiles}),
in line with the observations of \citet{Irwin1998,Irwin2001} and
\citet{sromovsky2002jupiter}. This invariance cannot be a consequence
of a constant temperature-pressure profile between the zones and the
belts, as we found that varying the temperature profile systematically
by 5 K, in line with measured zone-belt variations in temperature
in the upper troposphere \citep{fletcher2016mid}, produced no effect
on the retrieved cloud altitude. We also found no evidence of a discrete
cloud layer near the ammonia condensation level as found by \citet{banfield1998jupiter}
and \citet{simon2001color} in the visible and near-IR, or as required
in the mid-IR (eg. \citealp{matcheva2005cloud}). This cannot simply
be explained by effects of the model, as suppressing prior cloud densities
below the predicted ammonia cloud level, usually estimated around
0.7 bars (eg. \citet{atreya1999comparison}), leads to a very poor
fit to continuum wavelengths regardless of particle size distribution,
as we show clearly in Figure \ref{fig:ezcloudlevel}. Even in the
EZ, in which net upwelling is thought to be at its strongest, we always
required a thick cloud layer deeper than the 1-bar level to fit our
spectra. It is predicted to be too warm at these altitudes for ammonia
ice to condense out of the atmosphere, and so these cloud layers cannot
be made of ammonia ice. Although ECCMs usually predict an NH\textsubscript{4}SH
cloud to be located deeper than our retrieved altitude (around 2.2-2.6
bars as estimated eg. by \citet{atreya1999comparison}), the cloud
could be located at a higher altitude if the H\textsubscript{2}S
mixing ratio is suppressed, or if the cloud is made of a different
sulphur compound. However, this would conflict with our observations
of NEB cloud which appear to have a low real refractive index. We
are unable to reconcile these two conflicting observations without
speculation, perhaps invoking the presence of ammonia and water ice
`mushballs' that circulate around the main cloud layers as hypothesised
by \citet{2019EPSC...11..246T}. It is also possible that the apparently
missing NH\textsubscript{3} cloud is in fact present in the zones,
but cannot be resolved in our data either from the main 1.4-bar cloud
layer or from the upper tropospheric hazes due to a lack of vertical
resolution.

In spite of the invariance of the main cloud altitude, we generally
observed the zone-belt variations in colour, cloud structure and ammonia
abundance that we expected to find. The zones were associated with
thicker cloud, less colour, greater cloud-top ammonia abundances,
larger particle sizes and greater haze thickness than the belts. Upper
tropospheric haze was concentrated around the EZ, as it has been since
the year 2000 \citep{lii2010temporal}, and stratospheric haze thickest
and highest towards the poles. We found that the STropZ had cloud
opacities larger that those of the EZ, an observation that the reflectivity
measurements of \citet{Mendikoa2017} would also lend some credence
to. After the GRS, the southern NEB was the reddest region of Jupiter;
although the NTBs ws still red at this time, its colour was not as
prominent as it had been in 2017 as shown in Table \ref{tab:fitparameters}.
This shows that our model provides valid results on the properties
of Jupiter's atmosphere. However, we did obtain some results that
defied our expectations. One was the surprising lack of cloud cover
even in the southern part of the EZ that was relatively unaffected
by the colouration event that started around that time \citep{antunano2018infrared},
which was comparable in its thickness to that found in the northern
NEB due to mixing from the North Tropical Zone (NTropZ). Surprisingly
high chromophore abundances are retrieved in the northern half of
the EZ, even though the colouration event was still in its infancy
at this stage.%
{} We also note that the region where the red NTBs colour was still
prominent was barely distinct from the surrounding region in terms
of either cloud structure or gaseous ammonia abundances, which it
was in early 2017 when it was associated with thick cloud and haze
cover. This may hint at chromophore in this region that is located
in a stable region of the atmosphere unaffected by the weather below
it.

Unfortunately, chromophore abundances outside the GRS were insufficient
even in the NEB to completely constrain their altitude in the atmosphere,
as we lack the wavelengths below 475 nm that are sensitive to the
stratosphere. However, they were sufficient in the belts to constrain
a `lower bound' on their likely altitude, that is, the deepest pressure
level that the chromophore could be located at. This is usually around
(0.6$\pm$0.1) bars, well above the retrieved altitude of the deepest
visible cloud layer, but located at approximately the level of the
predicted ammonia ice cloud layer. We therefore cannot rule out zone-belt
differences in colour being due to the sublimation of ammonia ice
rime in the warmer belts to reveal chromophore nuclei.%

\section{Conclusions}

We analysed three spectral image cubes of Jupiter, obtained between
2014 and 2018 in the visible and near-infrared (480-930 nm) from the
VLT/MUSE instrument, in order to a) characterise the absorption spectrum
of Jupiter's colour-carrying compounds (`chromophores') and whether
it could be applied uniformly to model all blue-absorption on Jupiter
(a `universal chromophore'), and b) to analyse spatial variability
in colour and cloud/haze structure over Jupiter's surface. This was
performed using a three-cloud model, consisting of two conservatively-scattering
layers corresponding to deep cloud and high-altitude haze respectively,
and a single chromophore layer whose absorption spectrum was found
by directly retrieving the imaginary part of its refractive index.
Our main conclusions were as follows:
\begin{itemize}
\item As we were able to fit the shape of the shortwave spectral feature
of both the belts and the GRS well using the same chromophore compound,
we cannot rule out the possibility of a universal chromophore as proposed
by \citet{Sromovsky2017}. However, in order to fit the most recent
observations of the GRS following its recent intensification in colour,
we found that this chromophore compound required steeper blue-absorption
than could be provided by the tabulated optical constants of \citet{carlson2016chromophores}
in order to be in keeping with previous constraints on the opacity
of the main cloud layer. This is nonetheless in keeping with a chromophore
production mechanism involving photolysed ammonia reacting with acetylene
as proposed by \citet{Ferris1987,Ferris1988}, although the issue
with the predicted relative absence of acetylene that would be required
to produce chromophore in the troposphere according to this mechanism
remains to be resolved.
\item We identified some spectral variability between 510 and 540 nm that,
while well within the spectral error constraints provided by VLT/MUSE,
does alter the shape of the blue-absorption gradient found in spectra
throughout Jupiter, which we were unable to model through changes
in cloud particle size alone. This could potentially be due to a secondary
chromophore absorption feature, for which we recommend further investigation.
\item Although the exact vertical location of chromophore was difficult
to constrain given our wavelength range, we had sufficient information
to constrain a `lower bound' on how deep the chromophore could be
located. In the belts, we found that the chromophore could not be
located deeper than around $0.6\pm0.1$ bars, approximately equivalent
to the predicted location of the ammonia ice cloud layer on Jupiter.
In the GRS on the other hand, we found that the chromophore must be
located in or above the upper tropospheric haze, around 0.2 bars.
\item We retrieved a thick cloud layer whose base altitude varied little
from $1.4\pm0.3$ bars over the surface of Jupiter. Temperatures are
too high at these altitudes for ammonia to condense, and hence this
cloud layer could not be primarily made of ammonia ice. On the other
hand, we found that the retrieved cloud opacities and particle sizes
in the belts and the GRS were in keeping with a cloud layer that had
a low refractive index close to that of ammonia ice (1.42, \citet{martonchik1984optical}),
and could not be as high as 1.85 as was found by \citet{sato2013retrieval}.
This appears to contradict a cloud layer being made primarily of NH\textsubscript{4}SH
at these altitudes, whose refractive index is predicted to be substantially
larger than that of ammonia ice, and may be the result of overlying
ammonia ice masking the signature of the NH\textsubscript{4}SH cloud
layer.
\end{itemize}
Facility: VLT/MUSE

\section*{Acknowledgements}

This paper makes use of data from the European Southern Observatory
(ESO), programme IDs 60.A-9100, 099.C-0192 and 101.C-0097, as well
as data acquired from the NASA/ESA HST Space Telescope, associated
with OPAL program (PI: Simon, GO13937), and archived by the Space
Telescope Science Institute, which is operated by the Association
of Universities for Research in Astronomy, Inc., under NASA contract
NAS 5-26555. All maps of the latter are available at http://dx.doi.org/10.17909/T9G593.
We thank Larry Sromovsky for his codes to calculate Rayleigh-scattering
opacities and Mark Loeffler for providing absorption spectra of irradiated
NH\textsubscript{4}SH. Ashwin Braude was supported through a studentship
from the United Kingdom Science and Technology Facilities Council
(STFC). Glenn Orton was supported by funding from NASA distributed
to the Jet Propulsion Laboratory, California Institute of Technology.
Leigh Fletcher was supported by a Royal Society Research Fellowship
at the University of Leicester.

\bibliographystyle{plainnat}
\bibliography{C:/Users/braude/Desktop/Backup/jupiter_muse_paper/thesisbib}

\clearpage{}

\begin{table}
\begin{tabular*}{1\textwidth}{@{\extracolsep{\fill}}|c|c|c|c|}
\hline 
{\footnotesize{}Observation ID} & {\footnotesize{}Seeing (arcsec)} & {\footnotesize{}Airmass} & {\footnotesize{}Sub-observer longitude (III)}\tabularnewline
\hline 
\hline 
{\footnotesize{}2014-02-17T02:07:56.907} & {\footnotesize{}0.64} & {\footnotesize{}1.500} & {\footnotesize{}18.92}\tabularnewline
\hline 
{\footnotesize{}2017-05-15T02:01:59.328} & {\footnotesize{}1.10} & {\footnotesize{}1.067} & {\footnotesize{}165.5}\tabularnewline
\hline 
{\footnotesize{}2018-04-09T06:04:06.918} & {\footnotesize{}0.36} & {\footnotesize{}1.028} & {\footnotesize{}121.14}\tabularnewline
\hline 
\end{tabular*}

\caption{List of MUSE observations analysed in this work.}

\label{listofobs}
\end{table}

\begin{table}
\begin{adjustbox}{max totalsize={\textwidth}{\textheight}}%
\begin{tabular}{|>{\raggedright}p{0.3\textwidth}|>{\raggedright}p{0.4\textwidth}|l|l|>{\raggedright}p{0.3\textwidth}|}
\hline 
\textbf{Profile} & \textbf{Variable definition} & \textbf{Variable symbol} & \textbf{Fixed (F) or variable (V)?} & \textbf{Constraints where applicable}\tabularnewline
\hline 
\hline 
\multirow{9}{0.3\textwidth}{Cloud (P>0.15 bars), haze (P<0.15 bars)} & Aerosol abundance (optical depth/bar at 890 nm) as a function of pressure
$P$ (in bars) & \multirow{1}{*}{$\frac{\tau_{n}(P)}{\Delta p}$, $\frac{\tau_{h}(P)}{\Delta p}$} & \multirow{1}{*}{V} & \multirow{1}{0.3\textwidth}{}\tabularnewline
\cline{2-5} \cline{3-5} \cline{4-5} \cline{5-5} 
 & Correlation length of aerosol profile & $\Lambda_{\tau n}$, $\Lambda_{\tau h}$ & \multirow{6}{*}{F} & $1.5$\tabularnewline
\cline{2-3} \cline{3-3} \cline{5-5} 
 & Variance of particle size distribution & $\sigma_{n}$, $\sigma_{h}$ &  & $0.05$\tabularnewline
\cline{2-3} \cline{3-3} \cline{5-5} 
 & Real refractive index (cloud) at $\lambda=700nm$ & $n_{n}$ &  & Found through $\chi^{2}/n$ analysis\tabularnewline
\cline{2-3} \cline{3-3} \cline{5-5} 
 & Real refractive index (haze) at $\lambda=700nm$ & $n_{h}$ &  & $1.4$\tabularnewline
\cline{2-3} \cline{3-3} \cline{5-5} 
 & Imaginary part of complex refractive index spectrum as a function
of wavelength $\lambda$ & \multirow{1}{*}{$k_{n}(\lambda)$, $k_{h}(\lambda)$} &  & \multirow{1}{0.3\textwidth}{$10^{-9}\forall\lambda$}\tabularnewline
\cline{2-3} \cline{3-3} \cline{5-5} 
 & Correlation length of refractive index spectrum & $\Lambda_{kn}$, $\Lambda_{kh}$ &  & 0.1\tabularnewline
\cline{2-5} \cline{3-5} \cline{4-5} \cline{5-5} 
 & Effective cloud particle radius (\textgreek{m}m) & $r_{n}$ & V{*} & Found through $\chi^{2}/n$ analysis\tabularnewline
\cline{2-5} \cline{3-5} \cline{4-5} \cline{5-5} 
 & Effective haze particle radius (\textgreek{m}m) & $r_{h}$ & F{*} & $r_{h}<r_{n}$\tabularnewline
\hline 
\multirow{8}{0.3\textwidth}{Chromophore} & Aerosol abundance at centre of Gaussian (optical depth/bar at 890
nm) & $\frac{\tau_{c}(P)}{\Delta p}$ & \multirow{2}{*}{V} & \tabularnewline
\cline{2-3} \cline{3-3} \cline{5-5} 
 & Altitude of centre of Gaussian (bars) & $P_{c}$ &  & \tabularnewline
\cline{2-5} \cline{3-5} \cline{4-5} \cline{5-5} 
 & Gaussian FWHM (pressure scale height) & $\Delta_{c}$ & \multirow{5}{*}{F} & $0.25$\tabularnewline
\cline{2-3} \cline{3-3} \cline{5-5} 
 & Effective particle radius (\textgreek{m}m) & $r_{c}$ &  & Found through $\chi^{2}/n$ analysis\tabularnewline
\cline{2-3} \cline{3-3} \cline{5-5} 
 & Variance of particle size distribution & $\sigma_{c}$ &  & $0.1$\tabularnewline
\cline{2-3} \cline{3-3} \cline{5-5} 
 & Real refractive index at $\lambda=700nm$ & $n_{c}$ &  & $1.4$\tabularnewline
\cline{2-3} \cline{3-3} \cline{5-5} 
 & Correlation length of refractive index spectrum & $\Lambda_{kc}$ &  & $0.1$\tabularnewline
\cline{2-5} \cline{3-5} \cline{4-5} \cline{5-5} 
 & Imaginary part of complex refractive index spectrum as a function
of wavelength $\lambda$ & $k_{c}(\lambda)$ & F{*} & \tabularnewline
\hline 
\multirow{3}{0.3\textwidth}{Ammonia gas} & Reference pressure level (bars) & $nh_{3}p_{1}$ & F & 0.6\tabularnewline
\cline{2-5} \cline{3-5} \cline{4-5} \cline{5-5} 
 & Reference volume mixing ratio (VMR) at a pressure level of 0.6 bars & \multirow{1}{*}{$nh_{3}v_{1}$} & \multirow{1}{*}{V} & Saturation-limited\tabularnewline
\cline{2-5} \cline{3-5} \cline{4-5} \cline{5-5} 
 & \multirow{1}{0.4\textwidth}{Fractional scale height} & \multirow{1}{*}{fsh} & \multirow{1}{*}{V} & Saturation-limited, VMR must decrease with increasing altitude\tabularnewline
\hline 
\end{tabular}
\end{adjustbox}

\caption{\label{tab:Explanation-of-variables}Explanation of variables used
to parametrise the forward model, with prior constraints specified
when applicable. V{*} indicates variables that are fixed during limb-darkening
analysis, but allowed to vary when doing point spectral retrievals.
F{*} indicates variables that are allowed to vary during limb-darkening
analysis, but fixed when doing point spectral retrievals to the values
retrieved from limb-darkening.}

\label{carlsophoreparameters}
\end{table}

\begin{table}
\begin{adjustbox}{max width=\textwidth}

\begin{tabular}{|c|c|c|c|c|}
\hline 
$n_{n}$ & \multicolumn{3}{c|}{1.42} & 1.5\tabularnewline
\hline 
$r_{c}$ (\textgreek{m}m) & 0.05 & 0.2 & \multicolumn{2}{c|}{0.5}\tabularnewline
\hline 
$r_{n}$ (\textgreek{m}m) & \multicolumn{3}{c|}{1} & 0.75\tabularnewline
\hline 
Wavelength (nm) & \multicolumn{4}{c|}{$k_{c}$}\tabularnewline
\hline 
\hline 
450 & $0.033\pm0.003$ & $0.038\pm0.003$ & $0.048\pm0.005$ & $0.045\pm0.004$\tabularnewline
\hline 
500 & $0.023\pm0.001$ & $0.024\pm0.001$ & $0.027\pm0.001$ & $0.025\pm0.001$\tabularnewline
\hline 
550 & $0.016\pm0.001$ & $0.0154\pm0.0007$ & $0.0150\pm0.0007$ & $0.0151\pm0.0006$\tabularnewline
\hline 
600 & $0.0107\pm0.0007$ & $0.0098\pm0.0004$ & $0.0089\pm0.0004$ & $0.0098\pm0.0004$\tabularnewline
\hline 
650 & $0.0077\pm0.0005$ & $0.0069\pm0.0003$ & $0.0059\pm0.0003$ & $0.0070\pm0.0003$\tabularnewline
\hline 
700 & $0.0081\pm0.0005$ & $0.0075\pm0.0005$ & $0.0063\pm0.0003$ & $0.0074\pm0.0003$\tabularnewline
\hline 
750 & $0.0078\pm0.0005$ & $0.0072\pm0.0004$ & $0.0059\pm0.0003$ & $0.0068\pm0.0003$\tabularnewline
\hline 
800 & $0.0065\pm0.0005$ & $0.0057\pm0.0004$ & $0.0047\pm0.0003$ & $0.0048\pm0.0003$\tabularnewline
\hline 
850 & $0.0091\pm0.0007$ & $0.0088\pm0.0005$ & $0.0073\pm0.0004$ & $0.0052\pm0.0003$\tabularnewline
\hline 
900 & $0.0083\pm0.0007$ & $0.0079\pm0.0007$ & $0.0068\pm0.0005$ & $0.0041\pm0.0004$\tabularnewline
\hline 
950 & $0.007\pm0.001$ & $0.007\pm0.001$ & $0.0061\pm0.0008$ & $0.0047\pm0.0007$\tabularnewline
\hline 
\end{tabular}
\end{adjustbox}

\caption{Tabulated retrieved imaginary refractive index spectra as shown in
Figure \ref{fig:imagri}.}

\label{imagritable}
\end{table}

\begin{table}
\begin{adjustbox}{max width=\textwidth}{\footnotesize{}}%
\begin{tabular}{|l|c|c|c|c|c||c|c|>{\centering}p{0.12\textwidth}|>{\centering}p{0.14\textwidth}|>{\centering}p{0.18\textwidth}|>{\centering}p{0.19\textwidth}|>{\centering}p{0.18\textwidth}|>{\centering}p{0.18\textwidth}|}
\hline 
\multicolumn{6}{|c||}{{\footnotesize{}Retrieval from limb-darkening}} & \multicolumn{8}{c|}{{\footnotesize{}Application to GRS}}\tabularnewline
\hline 
{\footnotesize{}$\sigma$} & {\footnotesize{}$n_{n}$} & {\footnotesize{}$r_{c}$ (\textgreek{m}m)} & {\footnotesize{}$r_{n}$ (\textgreek{m}m)} & {\footnotesize{}$r_{h}$ (\textgreek{m}m)} & {\footnotesize{}$\chi^{2}/n$} & {\footnotesize{}$r_{n}$ (\textgreek{m}m)} & {\footnotesize{}$\chi^{2}/n$} & {\footnotesize{}Max. cloud+haze optical depth/bar} & {\footnotesize{}Max. chromophore optical depth/bar} & {\footnotesize{}Max. cloud+haze density (g/l)} & {\footnotesize{}Max. chromophore density (g/l)} & {\footnotesize{}Total cloud+haze column abundance (g/cm}\textsuperscript{{\footnotesize{}2}}{\footnotesize{})} & {\footnotesize{}Total chromophore column abundance (g/cm}\textsuperscript{{\footnotesize{}2}}{\footnotesize{})}\tabularnewline
\hline 
\hline 
\multirow{4}{*}{{\footnotesize{}1}} & \textbf{\footnotesize{}1.42} & \textbf{\footnotesize{}0.05} & \textbf{\footnotesize{}1} & \textbf{\footnotesize{}$\mathbf{0.47\pm0.03}$} & \textbf{\footnotesize{}1.12622} & \textbf{\footnotesize{}$\mathbf{6.58\pm0.05}$} & \textbf{\footnotesize{}0.663071} & \textbf{\footnotesize{}$\mathbf{11.2\pm0.7}$} & \textbf{\footnotesize{}$\mathbf{1.03\pm0.02}$\-} & \textbf{\footnotesize{}$\mathbf{(1.7\pm0.1)\times10^{-6}}$} & \textbf{\footnotesize{}$\mathbf{(6.6\pm0.2)\times10^{-8}}$} & {\footnotesize{}$\mathbf{(9\pm1)\times10^{-3}}$} & \textbf{\footnotesize{}$\mathbf{(6.3\pm0.3)\times10^{-5}}$}\tabularnewline
\cline{2-14} \cline{3-14} \cline{4-14} \cline{5-14} \cline{6-14} \cline{7-14} \cline{8-14} \cline{9-14} \cline{10-14} \cline{11-14} \cline{12-14} \cline{13-14} \cline{14-14} 
 & {\footnotesize{}1.42} & {\footnotesize{}0.2} & {\footnotesize{}1} & {\footnotesize{}$0.48\pm0.02$} & {\footnotesize{}1.14259} & {\footnotesize{}$7.62\pm0.01$} & {\footnotesize{}1.08430} & {\footnotesize{}$21\pm2$} & {\footnotesize{}$5.3\pm0.1$} & {\footnotesize{}$(5.9\pm0.6)\times10^{-6}$} & {\footnotesize{}$(3.15\pm0.07)\times10^{-8}$} & {\footnotesize{}$(2.8\pm0.6)\times10^{-2}$} & \textbf{\footnotesize{}$(3.1\pm0.1)\times10^{-5}$}\tabularnewline
\cline{2-14} \cline{3-14} \cline{4-14} \cline{5-14} \cline{6-14} \cline{7-14} \cline{8-14} \cline{9-14} \cline{10-14} \cline{11-14} \cline{12-14} \cline{13-14} \cline{14-14} 
 & {\footnotesize{}1.42} & {\footnotesize{}0.5} & {\footnotesize{}1} & {\footnotesize{}$0.48\pm0.02$} & {\footnotesize{}1.14848} & {\footnotesize{}$7.831\pm0.007$} & {\footnotesize{}2.37468} & {\footnotesize{}$49\pm6$} & {\footnotesize{}$13.0\pm0.5$} & {\footnotesize{}$(1.3\pm0.2)\times10^{-5}$} & {\footnotesize{}$(4.2\pm0.2)\times10^{-8}$} & {\footnotesize{}$(5\pm2)\times10^{-2}$} & \textbf{\footnotesize{}$(4.8\pm0.3)\times10^{-5}$}\tabularnewline
\cline{2-14} \cline{3-14} \cline{4-14} \cline{5-14} \cline{6-14} \cline{7-14} \cline{8-14} \cline{9-14} \cline{10-14} \cline{11-14} \cline{12-14} \cline{13-14} \cline{14-14} 
 & {\footnotesize{}1.5} & {\footnotesize{}0.5} & {\footnotesize{}0.75} & {\footnotesize{}$0.50\pm0.01$} & {\footnotesize{}1.16793} & {\footnotesize{}$0.094\pm0.004$} & {\footnotesize{}2.47952} & {\footnotesize{}$74\pm15$} & {\footnotesize{}$13.5\pm0.5$} & {\footnotesize{}$(8\pm2)\times10^{-6}$} & {\footnotesize{}$(2.5\pm0.9)\times10^{-8}$} & {\footnotesize{}$(3\pm1)\times10^{-2}$} & \textbf{\footnotesize{}$(2.3\pm0.1)\times10^{-5}$}\tabularnewline
\hline 
\hline 
\multirow{5}{*}{{\footnotesize{}2}} & {\footnotesize{}1.42} & {\footnotesize{}1} & {\footnotesize{}1} & {\footnotesize{}$0.45\pm0.03$} & {\footnotesize{}1.17243} & {\footnotesize{}$5.38\pm0.05$} & {\footnotesize{}1.43397} & {\footnotesize{}$430\pm70$} & {\footnotesize{}$15.6\pm0.4$} & {\footnotesize{}$(8\pm2)\times10^{-5}$} & {\footnotesize{}$(6.9\pm0.2)\times10^{-8}$} & {\footnotesize{}$(3\pm1)\times10^{-1}$} & \textbf{\footnotesize{}$(6.2\pm0.2)\times10^{-5}$}\tabularnewline
\cline{2-14} \cline{3-14} \cline{4-14} \cline{5-14} \cline{6-14} \cline{7-14} \cline{8-14} \cline{9-14} \cline{10-14} \cline{11-14} \cline{12-14} \cline{13-14} \cline{14-14} 
 & {\footnotesize{}1.6} & {\footnotesize{}0.02} & {\footnotesize{}0.5} & {\footnotesize{}$0.183\pm0.003$} & {\footnotesize{}1.20119} & {\footnotesize{}$0.89\pm0.01$} & {\footnotesize{}0.653765} & {\footnotesize{}$8.7\pm0.6$} & {\footnotesize{}$0.45\pm0.02$} & {\footnotesize{}$(8.3\pm0.8)\times10^{-8}$} & {\footnotesize{}$(5.4\pm0.2)\times10^{-8}$} & {\footnotesize{}$(5.2\pm0.8)\times10^{-4}$} & \textbf{\footnotesize{}$(5.2\pm0.3)\times10^{-5}$}\tabularnewline
\cline{2-14} \cline{3-14} \cline{4-14} \cline{5-14} \cline{6-14} \cline{7-14} \cline{8-14} \cline{9-14} \cline{10-14} \cline{11-14} \cline{12-14} \cline{13-14} \cline{14-14} 
 & {\footnotesize{}1.42} & {\footnotesize{}0.02} & {\footnotesize{}1} & {\footnotesize{}$0.52\pm0.03$} & {\footnotesize{}1.20469} & {\footnotesize{}$7.773\pm0.008$} & {\footnotesize{}1.06207} & {\footnotesize{}$11.5\pm0.9$} & {\footnotesize{}$1.5\pm0.1$} & {\footnotesize{}$(1.6\pm0.2)\times10^{-6}$} & {\footnotesize{}$(1.13\pm0.08)\times10^{-7}$} & {\footnotesize{}$(1.2\pm0.2)\times10^{-2}$} & \textbf{\footnotesize{}$(1.0\pm0.1)\times10^{-4}$}\tabularnewline
\cline{2-14} \cline{3-14} \cline{4-14} \cline{5-14} \cline{6-14} \cline{7-14} \cline{8-14} \cline{9-14} \cline{10-14} \cline{11-14} \cline{12-14} \cline{13-14} \cline{14-14} 
 & {\footnotesize{}1.6} & {\footnotesize{}0.5} & {\footnotesize{}0.5} & {\footnotesize{}$0.183\pm0.002$} & {\footnotesize{}1.2062} & {\footnotesize{}$1.07\pm0.03$} & {\footnotesize{}2.1343} & {\footnotesize{}$8.8\pm0.6$} & {\footnotesize{}$6.9\pm0.7$} & {\footnotesize{}$(2.7\pm0.2)\times10^{-7}$} & {\footnotesize{}$(3.3\pm0.3)\times10^{-8}$} & {\footnotesize{}$(1.0\pm0.1)\times10^{-3}$} & \textbf{\footnotesize{}$(3.3\pm0.6)\times10^{-5}$}\tabularnewline
\cline{2-14} \cline{3-14} \cline{4-14} \cline{5-14} \cline{6-14} \cline{7-14} \cline{8-14} \cline{9-14} \cline{10-14} \cline{11-14} \cline{12-14} \cline{13-14} \cline{14-14} 
 & {\footnotesize{}1.6} & {\footnotesize{}0.1} & {\footnotesize{}0.5} & {\footnotesize{}$0.40\pm0.06$} & {\footnotesize{}1.2149} & {\footnotesize{}$0.745\pm0.008$} & {\footnotesize{}0.850912} & {\footnotesize{}$15\pm1$} & {\footnotesize{}$1.91\pm0.05$} & {\footnotesize{}$(1.5\pm0.1)\times10^{-7}$} & {\footnotesize{}$(3.9\pm0.1)\times10^{-8}$} & {\footnotesize{}$(9\pm2)\times10^{-4}$} & \textbf{\footnotesize{}$(3.5\pm0.2)\times10^{-5}$}\tabularnewline
\hline 
\hline 
\multirow{7}{*}{{\footnotesize{}3}} & {\footnotesize{}1.42} & {\footnotesize{}0.05} & {\footnotesize{}0.75} & {\footnotesize{}$0.066\pm0.002$} & {\footnotesize{}1.22733} & {\footnotesize{}$4.76\pm0.04$} & {\footnotesize{}0.865846} & {\footnotesize{}$13.6\pm0.9$} & {\footnotesize{}$0.84\pm0.05$} & {\footnotesize{}$(1.0\pm0.1)\times10^{-6}$} & {\footnotesize{}$(4.5\pm0.3)\times10^{-8}$} & {\footnotesize{}$(6.5\pm0.9)\times10^{-3}$} & \textbf{\footnotesize{}$(4.4\pm0.4)\times10^{-5}$}\tabularnewline
\cline{2-14} \cline{3-14} \cline{4-14} \cline{5-14} \cline{6-14} \cline{7-14} \cline{8-14} \cline{9-14} \cline{10-14} \cline{11-14} \cline{12-14} \cline{13-14} \cline{14-14} 
 & {\footnotesize{}1.42} & {\footnotesize{}0.1} & {\footnotesize{}1} & {\footnotesize{}$0.40\pm0.05$} & {\footnotesize{}1.2375} & {\footnotesize{}$5.28\pm0.06$} & {\footnotesize{}0.751611} & {\footnotesize{}$11.4\pm0.7$} & {\footnotesize{}$1.84\pm0.05$} & {\footnotesize{}$(1.9\pm0.2)\times10^{-6}$} & {\footnotesize{}$(3.6\pm0.1)\times10^{-8}$} & {\footnotesize{}$(1.0\pm0.2)\times10^{-2}$} & \textbf{\footnotesize{}$(3.6\pm0.2)\times10^{-5}$}\tabularnewline
\cline{2-14} \cline{3-14} \cline{4-14} \cline{5-14} \cline{6-14} \cline{7-14} \cline{8-14} \cline{9-14} \cline{10-14} \cline{11-14} \cline{12-14} \cline{13-14} \cline{14-14} 
 & {\footnotesize{}1.42} & {\footnotesize{}0.02} & {\footnotesize{}0.75} & {\footnotesize{}$0.071\pm0.002$} & {\footnotesize{}1.23781} & {\footnotesize{}$6.86\pm0.04$} & {\footnotesize{}1.14264} & {\footnotesize{}$10.2\pm0.8$} & {\footnotesize{}$0.77\pm0.05$} & {\footnotesize{}$(7.4\pm0.6)\times10^{-7}$} & {\footnotesize{}$(5.3\pm0.3)\times10^{-8}$} & {\footnotesize{}$(6.0\pm0.8)\times10^{-3}$} & \textbf{\footnotesize{}$(4.9\pm0.4)\times10^{-5}$}\tabularnewline
\cline{2-14} \cline{3-14} \cline{4-14} \cline{5-14} \cline{6-14} \cline{7-14} \cline{8-14} \cline{9-14} \cline{10-14} \cline{11-14} \cline{12-14} \cline{13-14} \cline{14-14} 
 & {\footnotesize{}1.42} & {\footnotesize{}1} & {\footnotesize{}0.75} & {\footnotesize{}$0.071\pm0.002$} & {\footnotesize{}1.24142} & {\footnotesize{}$4.74\pm0.08$} & {\footnotesize{}1.24856} & {\footnotesize{}$46\pm6$} & {\footnotesize{}$12.4\pm0.5$} & {\footnotesize{}$(9\pm1)\times10^{-6}$} & {\footnotesize{}$(5.5\pm0.2)\times10^{-8}$} & {\footnotesize{}$(4\pm1)\times10^{-2}$} & \textbf{\footnotesize{}$(4.9\pm0.3)\times10^{-5}$}\tabularnewline
\cline{2-14} \cline{3-14} \cline{4-14} \cline{5-14} \cline{6-14} \cline{7-14} \cline{8-14} \cline{9-14} \cline{10-14} \cline{11-14} \cline{12-14} \cline{13-14} \cline{14-14} 
 & {\footnotesize{}1.6} & {\footnotesize{}0.05} & {\footnotesize{}0.5} & {\footnotesize{}$0.35\pm0.05$} & {\footnotesize{}1.24357} & {\footnotesize{}$0.87\pm0.02$} & {\footnotesize{}0.926087} & {\footnotesize{}$8.7\pm0.8$} & {\footnotesize{}$0.69\pm0.03$} & {\footnotesize{}$(6.2\pm0.8)\times10^{-8}$} & {\footnotesize{}$(5.2\pm0.2)\times10^{-8}$} & {\footnotesize{}$(6\pm1)\times10^{-4}$} & \textbf{\footnotesize{}$(4.7\pm0.4)\times10^{-5}$}\tabularnewline
\cline{2-14} \cline{3-14} \cline{4-14} \cline{5-14} \cline{6-14} \cline{7-14} \cline{8-14} \cline{9-14} \cline{10-14} \cline{11-14} \cline{12-14} \cline{13-14} \cline{14-14} 
 & {\footnotesize{}1.42} & {\footnotesize{}0.1} & {\footnotesize{}0.5} & {\footnotesize{}$0.065\pm0.002$} & {\footnotesize{}1.24681} & {\footnotesize{}$0.118\pm0.004$} & {\footnotesize{}0.918502} & {\footnotesize{}$16\pm2$} & {\footnotesize{}$2.4\pm0.1$} & {\footnotesize{}$(1.3\pm0.2)\times10^{-6}$} & {\footnotesize{}$(3.7\pm0.2)\times10^{-8}$} & {\footnotesize{}$(8\pm2)\times10^{-3}$} & \textbf{\footnotesize{}$(3.3\pm0.2)\times10^{-5}$}\tabularnewline
\cline{2-14} \cline{3-14} \cline{4-14} \cline{5-14} \cline{6-14} \cline{7-14} \cline{8-14} \cline{9-14} \cline{10-14} \cline{11-14} \cline{12-14} \cline{13-14} \cline{14-14} 
 & {\footnotesize{}1.5} & {\footnotesize{}0.1} & {\footnotesize{}0.75} & {\footnotesize{}$0.40\pm0.02$} & {\footnotesize{}1.24714} & {\footnotesize{}$1.14\pm0.02$} & {\footnotesize{}2.78717} & {\footnotesize{}$45\pm5$} & {\footnotesize{}$1.68\pm0.06$} & {\footnotesize{}$(1.5\pm0.1)\times10^{-6}$} & {\footnotesize{}$(3.5\pm0.1)\times10^{-8}$} & {\footnotesize{}$(4.4\pm0.8)\times10^{-3}$} & \textbf{\footnotesize{}$(3.5\pm0.2)\times10^{-5}$}\tabularnewline
\hline 
\end{tabular}
\end{adjustbox}

\caption{\label{tab:grssolutions}Results of the retrieval and selection of
our universal chromophore solution. For each row, the first six columns
show the results of fitting to the NEB through limb-darkening for
given values of $n_{n}$, $r_{n}$ and $r_{c}$, while the last eight
columns then show the results when the solution for $k_{c}(\lambda)$
retrieved through limb-darkening is applied to the spectrum of the
GRS from 2018 and $r_{n}$ allowed to vary from its prior value (but
$n_{n}$ and $r_{c}$ remain fixed). The rows are sorted in ascending
order of $\chi^{2}/n$, and within sets of 1- 2- and 3$\sigma$ of
the lowest $\chi^{2}/n$ from limb-darkening analysis (where $\sigma=\sqrt{1/500}\approx0.045$).
For the GRS retrievals, $n=435$ and so $\sigma=\sqrt{2/435}\approx0.068$.
The selected universal chromophore solution is highlighted in bold,
due to its low $\chi^{2}/n$ value for both the NEB and the GRS.}
\label{retrievalresults}
\end{table}

\begin{table}
\begin{adjustbox}{max width=\textwidth}%
\begin{tabular}{|l||l|l|l|l|l|l|l|}
\hline 
 & \textbf{EZ} & \textbf{NEB} & \textbf{NTBs (2017)} & \textbf{GRS (2000), unnormalised} & \textbf{GRS (2000), normalised} & \textbf{GRS (2014)} & \textbf{GRS (2018)}\tabularnewline
\hline 
\hline 
$\tau_{n}$($P_{max}$)/$\Delta p$ & $10.6\pm0.6$ & $6.1\pm0.5$ & $8.0\pm0.5$ & $3.7\pm0.4$ & $7.3\pm0.8$ & $12\pm2$ & $11.2\pm0.7$\tabularnewline
\hline 
$P_{max}$ & $1.4\pm0.1$ & $1.07\pm0.08$ & $1.4\pm0.1$ & $1.4\pm0.1$ & $1.4\pm0.1$ & $1.4\pm0.1$ & $0.23\pm0.02$\tabularnewline
\hline 
$\tau_{c}$/$\Delta p$ & $0.20\pm0.07$ & $0.22\pm0.02$ & $0.26\pm0.06$ & $0.16\pm0.05$ & $0.13\pm0.02$ & $0.54\pm0.05$ & $1.03\pm0.02$\tabularnewline
\hline 
$P_{c}$ & $0.30\pm0.02$ & $0.61\pm0.04$ & $0.35\pm0.03$ & $0.93\pm0.07$ & $0.7\pm0.1$ & $0.23\pm0.02$ & $0.19\pm0.02$\tabularnewline
\hline 
$r_{n}$ & $4.40\pm0.05$ & $1.5\pm0.2$ & $4.90\pm0.03$ & $1.3\pm0.2$ & $1.5\pm0.4$ & $1.8\pm0.3$ & $6.58\pm0.04$\tabularnewline
\hline 
$\chi^{2}/n$ & $0.85\pm0.07$ & $0.54\pm0.07$ & $0.56\pm0.07$ & $1.7\pm0.2$ & $1.9\pm0.2$ & $0.40\pm0.07$ & $0.66\pm0.07$\tabularnewline
\hline 
\end{tabular}
\end{adjustbox}

\caption{\label{tab:fitparameters}Retrieved parameters associated with the
locations whose fits are shown in Figures \ref{fig:grsspecfits} and
\ref{fig:fitcomparison}. $P_{max}$ is the pressure level (in bars)
at which the cloud optical depth per bar is greatest. For the GRS
(2000) case, the number of wavelengths is 63 and for the other spectra
it is 435.}
\label{specfitstable}
\end{table}

\end{document}